\documentclass{SciPost}

\hypersetup{
    colorlinks,
    linkcolor={red!50!black},
    citecolor={blue!50!black},
    urlcolor={blue!80!black}
}

\usepackage[bitstream-charter]{mathdesign}
\usepackage{setspace}
\usepackage{physics}
\usepackage{dsfont}
\usepackage{tensor}
\usepackage[normalem]{ulem}
\usepackage{xcolor}
\usepackage{graphicx}
\usepackage[export]{adjustbox}
\usepackage{mathtools}
\usepackage{mathbbol}
\usepackage[utf8]{inputenc}
\usepackage{amsmath}
\usepackage{bbold}
\usepackage{breqn}

\usepackage[english]{babel}
\usepackage{ragged2e}
\usepackage{etoolbox}
\usepackage{lipsum}
\usepackage{multirow}
\usepackage{tcolorbox}
\usepackage{latexsym}
\usepackage{enumitem}
\usepackage{titlesec}
\usepackage{url}
\usepackage{caption}
\usepackage{subcaption}
\usepackage{comment}
\usepackage{booktabs} 

\colorlet{darkgreen}{green!50!black}

\newtheorem{theorem}{Theorem}[section]

\newtheorem{remark}{Remark}[section]
\newtheorem{proposition}{Proposition}[section]
\newtheorem{lemma}{Lemma}[section]
\newtheorem{corollary}{Corollary}[section]
\newtheorem{definition}{Definition}[section]
\def\br{\begin{remark}\rm\small}
	\def\er{\end{remark}}
\def\bt{\begin{theorem}}
	\def\et{\end{theorem}}
\def\bd{\begin{definition}}
	\def\ed{\end{definition}}
\def\bp{\begin{proposition}}
	\def\ep{\end{proposition}}
\def\bl{\begin{lemma}}
	\def\el{\end{lemma}}
\def\bc{\begin{corollary}}
	\def\ec{\end{corollary}}
\def\beaq{\begin{eqnarray}}
	\def\eeaq{\end{eqnarray}}

\newcommand{\be}{\begin{equation}}
	\newcommand{\ee}{\end{equation}}
\newcommand{\beq}{\begin{equation}}
	\newcommand{\eeq}{\end{equation}}
\newcommand{\bea}{\begin{eqnarray}}
	\newcommand{\eea}{\end{eqnarray}}

\makeatletter
\newsavebox{\@brx}
\newcommand{\llangle}[1][]{\savebox{\@brx}{\(\m@th{#1\langle}\)}%
  \mathopen{\copy\@brx\kern-0.5\wd\@brx\usebox{\@brx}}}
\newcommand{\rrangle}[1][]{\savebox{\@brx}{\(\m@th{#1\rangle}\)}%
  \mathclose{\copy\@brx\kern-0.5\wd\@brx\usebox{\@brx}}}
\makeatother

\DeclareSymbolFont{usualmathcal}{OMS}{cmsy}{m}{n}
\DeclareSymbolFontAlphabet{\mathcal}{usualmathcal}

\fancypagestyle{SPstyle}{
\fancyhf{}
\lhead{\colorbox{scipostblue}{\bf \color{white} ~SciPost Physics Lecture Notes }}
\rhead{{\bf \color{scipostdeepblue} ~Submission }}

\fancyfoot[C]{\textbf{\thepage}}
}

\begin{document}

\pagestyle{SPstyle}

\begin{center}{\Large \textbf{\color{scipostdeepblue}{
A Background-Independent Closed String Action at Tree Level\\
}}}\end{center}

\begin{center}\textbf{
Amr A. Ahmadain\textsuperscript{1$\star$},
Alexander Frenkel\textsuperscript{2 \#} and
Aron C. Wall\textsuperscript{3$\dagger$}
}\end{center}

\begin{center}
{\bf 1} Department of Physics, Swansea University, Swansea, SA2 8PP, UK
\\
{\bf 2} Leinweber Institute for Theoretical Physics at Stanford, 382 Via Pueblo, Stanford, CA
94305, USA
\\
{\bf 3} Department of Applied  Mathematics and Theoretical Physics, University of Cambridge,\\Wilberforce Road, Cambridge, CB3 0WA, United Kingdom
\\[\baselineskip]
$\star$ \href{mailto:amrahmadain@gmail.com}{\small amrahmadain@gmail.com}\,,\quad
$\#$ \href{mailto:frenkelalexander1@gmail.com}{\small frenkelalexander1@gmail.com}\,,\quad
$\dagger$ \href{mailto:aroncwall@gmail.com}{\small aroncwall@gmail.com}
\end{center}

\section*{\color{scipostdeepblue}{Abstract}}
\textbf{\boldmath{%
We propose an off-shell bosonic string action that removes the renormalization window constraint of \cite{Ahmadain:2022tew}. To all orders in conformal perturbation theory, we allow for deformations of the worldsheet theory by any primary or descendant irrelevant deformation. Non-perturbatively, there are no spurious solutions on the space of all worldsheet theories with a unitary matter sector that flows from a UV fixed point. As part of our investigation of this action, we find non-smooth behavior in the Zamolodchikov $C$-function. Our results mostly apply to Euclidean-signature target spaces, but can be extended to Lorentzian backgrounds which are invariant under time translations and CTO symmetry.
}}

\vspace{\baselineskip}

\noindent\textcolor{white!90!black}{%
\fbox{\parbox{0.975\linewidth}{%
\textcolor{white!40!black}{\begin{tabular}{lr}%
  \begin{minipage}{0.6\textwidth}%
    {\small Copyright attribution to authors. \newline
    This work is a submission to SciPost Physics Lecture Notes. \newline
    License information to appear upon publication. \newline
    Publication information to appear upon publication.}
  \end{minipage} & \begin{minipage}{0.4\textwidth}
    {\small Received Date \newline Accepted Date \newline Published Date}%
  \end{minipage}
\end{tabular}}
}}
}


\vspace{10pt}
\noindent\rule{\textwidth}{1pt}
\tableofcontents
\noindent\rule{\textwidth}{1pt}
\vspace{10pt}


\section{Introduction}\label{sec:Introduction}
\label{sec:intro}
A non-perturbative, background-independent formulation of string theory would be an important step toward recovering semiclassical quantum gravity. By background-independence, we mean a string action formulated without reference to a particular target space background. In this paper, we continue a long-running program \cite{Friedan1980,Lovelace1984,SenHeteroticPRL1985,FT1,FT2,FT3,Witten-Hull:susy-heterotic:1985,Callan:1985ia,Lovelace:1986kr,Callan:1986ja,Curci:1986hi,TseytlinAnomaly1986,TseytlinCentralCharge1987,TseytlinRenormalizationMobius1988} to develop a worldsheet description\footnote{As opposed to something second quantized, such as String Field Theory.} that treats all on-shell and off-shell target space configurations on a priori equal footing. It is this worldsheet-centric approach that we refer to as off-shell string theory.

In the context of worldsheet string theory, target space configurations are (at least perturbatively in $g_s$) in one-to-one correspondence with 2d QFT worldsheet theories. This relationship is made clearest in the language of non-linear sigma models (NLSMs) -- field theories whose coupling constants are themselves functions of the fields. The classic example is the worldsheet theory of a string propagating in curved spacetimes:
\begin{equation}\label{eqn:NLSM-act}
    S_{w.s.} = \frac{1}{4\pi \alpha'}\int \sqrt{g(z)}d^2z\, \left(\left(g^{ab}G_{\mu \nu}(X^{\mu}) + i\epsilon^{ab}B^{\mu \nu}(x^{\mu})\right)\partial_aX^{\mu} \partial_b X^{\nu} + \alpha' R\Phi(X^{\mu})\ldots \right) + S_g.
\end{equation}
$G_{\mu \nu}(X^{\mu})$ is the target space metric, and $\Phi(X^{\mu})$ is the target space dilaton. $S_g$ is the $bc$ ghost action. More generally, both supergravity fields and massive string modes will source additional couplings in the worldsheet action. 

\begin{figure}[h]
\centering
\includegraphics[width=\textwidth]{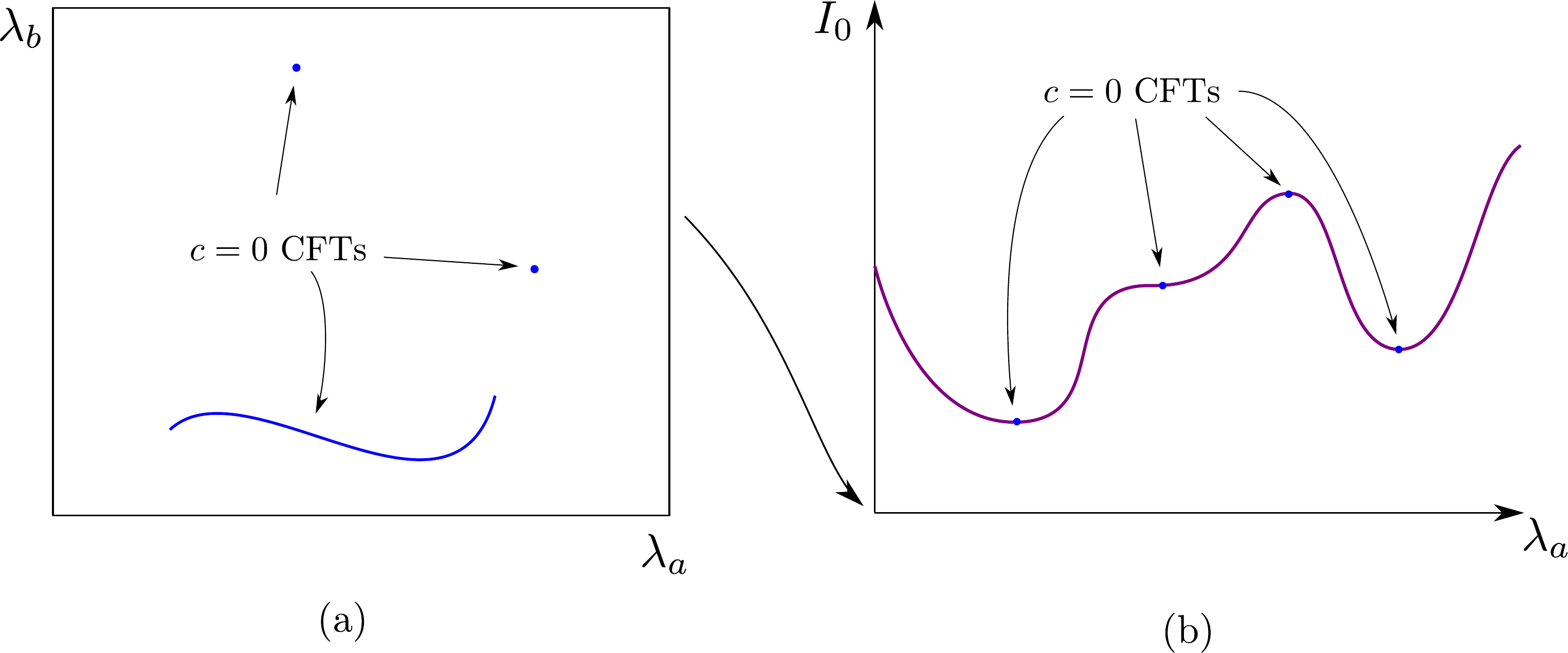}
\caption{(a) is a schematic depiction of the space of 2d QFTs. The dimensions of the space are labeled by worldsheet couplings (i.e. target space fields) $\lambda_{a}$. The marked points are string backgrounds, i.e. $c=0$ CFTs. On the space of QFTs, $c=0$ CFTs may either be isolated (depicted by the isolated points in the diagram) or appear as a continuous family (depicted by the curve). In (b), we have drawn a sketch of a tree-level string effective action $I_0$. $I_0$ is a function on this space of coupling constants. In order for $I_0$ to be a valid string effective action, its set of stationary points must precisely match the set of $c=0$ CFTs (or, equivalently, matter CFTs with $c = 26$). The purpose of this paper is to construct such an $I_0$.}\label{fig:qft-space}
\end{figure}

The space of 2d QFTs, therefore, must somehow take the place of the space of field configurations. The point of view we take is that the string effective action $I$ is a function on this space of 2d QFTs (see Fig. \ref{fig:qft-space}). At the classical level (i.e. tree level, the focus of this paper), our task is to find a function on this space whose fixed points are precisely classical string backgrounds -- $c=0$ CFTs. As developed in \cite{FT1,FT2,FT3} and generalized in \cite{Ahmadain:2022eso,Ahmadain:2022tew}, previous results have constructed valid effective actions to all orders in perturbation theory around string backgrounds. This paper extends these results non-perturbatively\footnote{It may be confusing about what it means to work non-perturbatively at tree level. A crisp discussion of this point is found in \cite{Ahmadain:2022tew}. Essentially, we work non-perturbatively in both $\alpha'$ and the number of external diagram legs $n$, while keeping to the sphere topology.} on the space of worldsheet theories with a unitary compact matter sector.

String theory is a theory of induced gravity \cite{Sakharov:1967pk,Visser:2002ew}\footnote{In string theory, target space has no dynamics of its own. The Einstein-Hilbert action and its coupling to matter is entirely induced by the string partition function.}, so the ansatz we take is that the classical effective action must somehow be computed from the worldsheet path integral. In particular, as we take $\hbar \rightarrow 0$ in target space (equivalently, $G_N \rightarrow 0$), the tree-level effective action must be computed from the sphere partition function $Z_{S^2}$. However, due to the noncompact SL$(2,\mathbb{C})$ conformal Killing group on the sphere, the string sphere partition function is a topic rife with subtleties \cite{LiuPolchinski1988,TSEYTLINMobiusInfinitySubtraction1988} (see also \cite{Maldacena:2001km,Kraus:2002cb,Troost:2011ud,Mahajan:2021nsd,Erler:2022agw,Eberhardt:2023lwd} for recent work on the subject in $\text{AdS}_3$ and string field theory). Along with the principles of renormalization group (RG) flow, these aspects form the foundations of Tseytlin's off-shell formalism.

We briefly introduce important known results. Consider the sphere partition function of the matter sector of the NLSM in \eqref{eqn:NLSM-act} \cite{TseytlinZeroMode1989,Ahmadain:2022eso}:
\begin{equation}\label{eqn:NLSM-Z0}
Z_{S^2} = \int d^DY\,\sqrt{G(Y)}e^{-2\Phi(Y)}\left[1 - \alpha' \log \epsilon \left(R + 2 \nabla^2 \Phi(Y)\right)\right] + O(\alpha{'}^2).
\end{equation}
For a $D$-dimensional target space, $Z_{S^2}$ is expressed as an integral over the $D$ worldsheet zero modes $Y^{\mu}$. $\epsilon$ is a UV cutoff introduced to make the partition function finite, and $R$ is the target space Ricci scalar. Notice that not only does the partition function seem to have a generally covariant form, but the $\alpha' \log \epsilon $ corrections to the worldsheet partition function precisely have the form of the target space string-frame Einstein-Hilbert action. In a series of papers \cite{FT1, FT2, FT3, TseytlinAnomaly1986,TseytlinWeylInvarianeCond1987,TseytlinRenormalizationMobius1988,TSEYTLINMobiusInfinitySubtraction1988,TseytlinSigmaModelEATachyons2001} Fradkin and Tseytlin built off of this observation to propose two concrete prescriptions for deriving the sphere effective action from the worldsheet:
\begin{equation}\label{eqn:tseyt-acts}
\begin{split}
&I_0^{\textbf{T1}} := -\frac{\partial}{\partial \log \epsilon}Z_{S^2},\\
&I_0^{\textbf{T2}} := -\frac{\partial}{\partial \log \epsilon}\left(1 + \frac{1}{2}\frac{\partial}{\partial \log \epsilon}\right)Z_{S^2}.
\end{split}
\end{equation}
Note that we must choose a particular Weyl frame $\omega(z)$ on which to evaluate $Z_{S^2}$. It may appear that this breaks local worldsheet conformal invariance, which must be a gauge redundancy in string theory. However, it was shown in \cite{Ahmadain:2022tew} (and extended to greater generality in \S\ref{ssec:field-redef}) that changes in the choice of $\omega(z)$ on the worldsheet are equivalent to field redefinitions in target space. This is directly related to local RG flow of the QFT. As a result, physical quantities calculated in target space are in fact independent of the choice of $\omega(z)$, and gauge invariance, crucially, remains unbroken.

In the context of \eqref{eqn:NLSM-Z0}, it is obvious why $I_0^{\textbf{T1}}$ and $I_0^{\textbf{T2}}$ work as effective actions to leading order in $\alpha'$. It is far less obvious why these prescriptions are not ad-hoc and continue to work to all orders in conformal perturbation theory for marginal and relevant worldsheet deformations\footnote{Corresponding to massless and tachyonic target space modes.} respectively. The crux of the physics is that the logarithmic derivatives in \eqref{eqn:tseyt-acts} are induced by modding out by the regularized (and gauge-orbit dependent) volume of SL$(2,\mathbb{C})$ \cite{LiuPolchinski1988,TseytlinRenormalizationMobius1988,TSEYTLINMobiusInfinitySubtraction1988,Ahmadain:2022tew}. Clarifying this relationship is central to the work taken up in \cite{Ahmadain:2022tew}. However, there is a more practical approach to understanding why these actions work and their limitations -- the language of spurious tadpoles. We now briefly review this point, as it is from this perspective that will construct an action that expands the regime of validity of the off-shell formalism non-perturbatively far from $c=0$ CFTs for all possible worldsheet deformations.

It is instructive to ask why $Z_{S^2}$ itself, before the application of any differential operators, fails as a target space effective action. The reason is that it is not stationary at $c=0$ CFTs. To see this, consider some coupling constant $\lambda^a$ parameterizing the space of 2d QFTs as depicted in Fig. \ref{fig:qft-space}(a). Derivatives with respect to a coupling constant are equivalent to inserting the operator it sources:
\begin{equation}
\frac{\partial Z_{S^2}}{\partial \lambda^a} = \int_{S^2}\sqrt{g(z)}d^2z\,\langle \langle \mathcal{O}_a(z) \rangle \rangle.
\end{equation}
As in \cite{Ahmadain:2022tew}, we use a double angle bracket $\langle \langle \cdot \rangle \rangle$ to denote an unnormalized expectation value, where we do not divide by the partition function $Z_{S^2}$. At a $c=0$ CFT all of these derivatives must vanish in order for it to be a stationary point. Indeed almost all of them do, as for any nontrivial operator conformal invariance demands that its one-point function vanish\footnote{This is true for a compact CFT -- the case we restrict ourselves to in this paper. For subtleties regarding non-compact CFTs see \cite{Ahmadain:2022tew,Kraus:noncompactCFT:2002}.}. We refer to these one-point functions as tadpoles. However, there is a family of operators for which the tadpole does not vanish in a $c=0$ CFT: the identity operator coupled to powers of the Ricci curvature -- $\mathcal{O}(z) = R^p \mathbb{1}$. The operator $R^p\mathbb{1}$ has scaling dimension $\Delta_p = 2p-2$. Call the corresponding coupling of this operator $t^{(p)}$, as in \cite{Ahmadain:2022tew}. We may directly calculate
\begin{equation}\label{eqn:spur-tad}
\begin{split}
&\left.\frac{\partial Z_{S^2}}{\partial t^{(p)}}\right|_{c=0\text{ CFT}} = \epsilon^{2p-2}Z_{S^2}^{(0)}\int_{S^2}d^2z\,\sqrt{g(z)}R^p(z)\\
&\iff Z_{S^2} = \ldots + t^{(p)}\epsilon^{2p-2}Z_{S^2}^{(0)}\int_{S^2}d^2z\,\sqrt{g(z)}R^p(z) + \ldots,
\end{split}
\end{equation}
where $Z_{S^2}^{(0)}$ is the unperturbed sphere partition function\footnote{Or equivalently, the generalized volume of target space.} coming from evaluating the unnormalized expectation value $\langle \langle \mathbb{1}\rangle\rangle$. For convenience, in order to make the coupling $t^{(p)}$ unitless we have explicitly factored out the dependence on the UV cutoff $\epsilon$. We may now explicitly see from \eqref{eqn:spur-tad} that $Z_{S^2}$ has linear dependence on a class of couplings $t^{(p)}$ parameterizing the space around a $c=0$ CFT. By definition, it fails to be a stationary point, so $Z_{S^2}$ fails to be a valid effective action.

Observe from \eqref{eqn:tseyt-acts} that the differential operator defining $I_0^{\textbf{T1}}$ has $\epsilon^0$ in its kernel, and the differential operator defining $I_0^{\textbf{T2}}$ has both $\epsilon^0$ and $\epsilon^{-2}$ in its kernel. These are precisely the $\epsilon$ dependence of the $p=1$ and $p=0$ terms\footnote{Referred to as the dilaton and tachyon tadpoles respectively, as these are precisely the vertex operators for the target space dilaton and tachyon.} in \eqref{eqn:spur-tad}. These operators therefore remove these linear terms from $Z_{S^2}$, such that
\begin{equation}
\frac{\partial I_0^{\textbf{T1}}}{\partial t^{(1)}} = 0, \quad \frac{\partial I_0^{\textbf{T2}}}{\partial t^{(0)}} = \frac{\partial I_0^{\textbf{T2}}}{\partial t^{(1)}} = 0.
\end{equation}
It is precisely this behavior that sets the regimes of validity of the \textbf{T1} and \textbf{T2} prescriptions. $I_0^{\textbf{T1}}$ and $I_0^{\textbf{T2}}$ are only valid target space effective actions on a subset of worldsheet deformations constrained by `renormalization windows' (in the language of \cite{Ahmadain:2022tew}). At order $n$ in conformal perturbation theory, these renormalization windows set bounds on the conformal dimension $\Delta = h + \bar{h}$ of the worldsheet deformations we allow:
\bea \label{eq:RW}
&{\bf T1}:&\quad \:2 - 2/n < \Delta < 2 + 2/n, \\
&{\bf T2}:&\quad \phantom{i2-2/}0 \le \Delta < 2 + 2/n.
\eea
The lower bound on the renormalization window of \textbf{T2} is set by unitarity.

An important role is played by couplings dressed with more than one power of $R(z)$. We refer to these as `higher non-minimal couplings'. In particular, spurious tadpoles with $p \geq 2$ fall into this category. It is these higher tadpoles that are not removed by the prescriptions \eqref{eqn:tseyt-acts}, and therefore set the limits on their regime of validity. We distinguish higher non-minimal couplings from `quasi-minimal' couplings -- those dressed with at most one power of the worldsheet Ricci scalar. In \S\ref{ssec:high-nonm}, we show that quasi-minimal couplings are the only physical degrees of freedom in the action we study in this paper -- \textit{all} higher non-minimal couplings, not just those responsible for spurious tadpoles, vanish from the action and equations of motion. They are therefore interpreted as pure gauge.

It is important to emphasize that the approach considered in this paper is distinct from that of string field theory (SFT) \cite{Erler-SFTLectureNotes-:2019,Erbin:2021,Maccaferri:2023vns,Sen:2024nfd}, another promising candidate for a background-independent formulation of string theory. The canonical formulation of the closed SFT action \cite{Zwiebach:1992ie} is perturbation around CFTs, where the classical solutions of the action is the space of local operators in the reference CFT. It is only within that space that local background independence was proved \cite{Sen:1990hh,Sen:1990na,Sen:1992pw,Sen:1993mh,Sen:1993kb}.
In \cite{Zwiebach:1996ph,Zwiebach:1996jc}, a systematic attempt to build a closed SFT action around (arbitrary) nonconformal backgrounds was considered within the Batalin-Vilkovisky (BV) formalism \cite{Batalin:1983ggl}, but the formulation is far from  being complete (See section 10 in \cite{Sen:2024nfd}). It would be interesting, in the future, to explore the relationship between the SFT action and our result.


\subsection{Overview of Results}

Previous attempts to include massive deformations \cite{Lovelace:1986kr, Banks:1987qs,Hughes:1988bw,Brustein:1987qw,Brustein:1991py} using the exact renormalization group equation 
\cite{Wilson:1973,Polchinski:ExactRG:1983} faced two main challenges: (1) renormalizing the spacetime action -- i.e. making it independent of the UV cutoff (RG scale) of the 2d QFT when very large dimension couplings are included, and (2) meeting the requirement of gauge invariance. In this work we make some progress on (1), but since we do not use the BRST formalism, it is unclear how the gauge invariance we find in this work relates to the usual notion of target space gauge invariance in string theory. We leave this for future analysis.

We additionally leave for future work supersymmetry, open strings, general Lorentzian target spaces, and higher genus (string loop) corrections. In this paper, we restrict our focus to unitary $c=26$ matter sectors (leaving the $c=-26$ ghost sector untouched) and the tree-level effective action. It is in this setting that we will be able to go non-perturbatively far away from $c=0$ CFTs. Furthermore, for simplicity we restrict our attention to compact target spaces, so the set of conformal dimensions of operators in the theory is discrete\footnote{As per the discussion in \cite{Ahmadain:2022tew}, we expect our results apply equally well to normalizable deformations of noncompact target spaces.}. With these caveats in mind, the main result of this paper is a sphere effective action valid over the entire space of unitary 2d QFTs. 

\begin{figure}[ht]
\centering
\includegraphics[width=\textwidth]{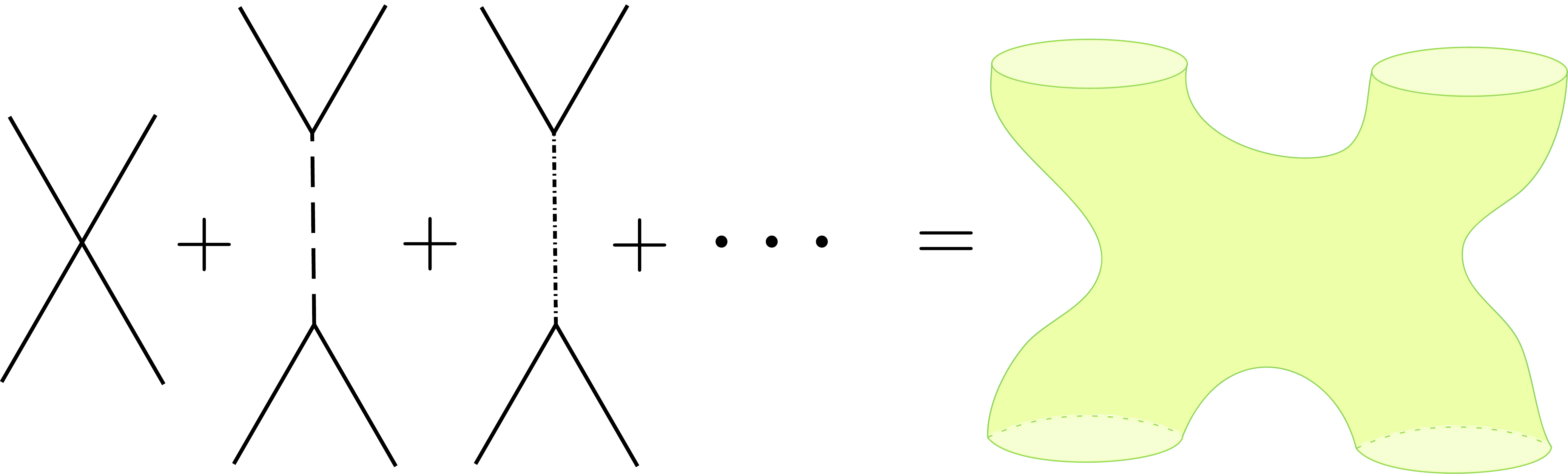}
\caption{We emphasize that the sphere effective action we compute from the worldsheet has already summed all contributions of all possible tree-level diagrams, including all possible massive species that may form an internal propagator. For example, in this figure we consider the four-point scattering amplitude of some massless field target space field $\phi$ (represented by the solid lines). The coefficient of the $\phi^4$ term in the effective action we compute is the sum of the $s,t,u$ channels (summed over all possible internal modes, represented by the different styles of dashed lines) and the 4-point contact interaction. This is contrary to the assumption made in \cite{Tseytlin-SFTComponents-1987}, which argues for a no-go theorem under the assumption that the target space effective action generates the 1PI diagrams at tree-level, not the full tree-level amplitude.}\label{fig:res-diag}
\end{figure}

In the language of \cite{Ahmadain:2022tew}, we remove any renormalization window on worldsheet deformations. We allow all possible irrelevant deformations, including descendants. Therefore, unlike the actions in \eqref{eqn:tseyt-acts} the action we consider \S\ref{sec:non-pert} includes string modes of arbitrarily large mass. It is thus important to ask how our results relate to \cite{Tseytlin-SFTComponents-1987}, which argues that it is impossible to write down a generally covariant action that reproduces the string S-matrix for all massless and massive modes. The argument relies on the observation that the three-point vertex in the string field theory action already reproduces the correct S-matrix for the scattering of massless modes, so any higher-point interactions appearing in the target space effective action must conspire to cancel at tree level. The important distinction between \cite{Tseytlin-SFTComponents-1987} and our perspective is depicted in Fig. \ref{fig:res-diag}. For us, the string effective action generates the entire $n$-point scattering amplitude -- not just the 1PI contribution that is typical of effective actions in QFT. Importantly, the action we write down still reproduces the correct equations of motion despite this difference. We leave a careful analysis of this aspect in asymptotically flat spaces with an S matrix for future work, as a completely satisfactory analysis requires a better understanding of noncompact target spaces.

\begin{figure}[h]
\centering
\includegraphics[width=0.5\textwidth]{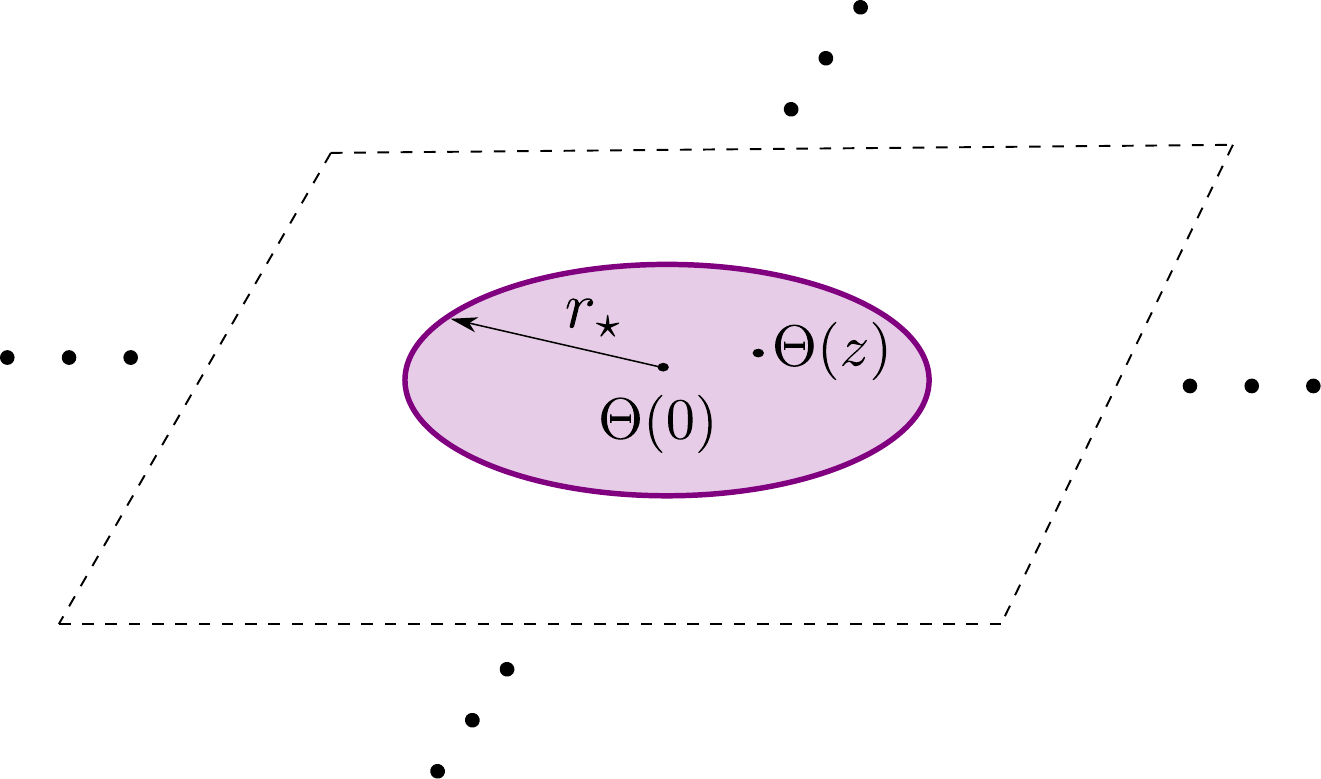}
\caption{A pictorial representation of the planar c-function as defined in \eqref{eqn:int-c}. The region drawn is a subset of the infinite plane. The shaded circle of radius $r_{\star}$ is the region of integration of $|z|^2\langle \Theta(z) \Theta(0) \rangle$ While $\Theta(0)$ is fixed at the origin, $\Theta(z)$ varies over a disk of radius $r_{\star}$.}\label{fig:planar-c-func}
\end{figure}

Our approach builds on an intriguing connection between $I_0^{\textbf{T2}}$ and the Zamolodchikov $C$-theorem \cite{Zamolodchikov1986}. The main observation is that the \textbf{T2} action can be expressed as the two-point function of the trace of the stress-energy tensor integrated over the sphere of arbitrary radius $r$:
\bea\label{eq:T2action-2pt-trace}
I_0^{\bf T2} =-\int d^2z \frac{|z|^2}{1 + |z|^2} \llangle \Theta(0) \Theta(z)\rrangle_{\text{QFT}},
\eea
where ${|z|^2}/({1+|z|^2}) = \tfrac{1}{2}(1 - \cos \theta)$. The potential connection to the planar $C$-theorem, which we review in section \ref{sec:review}, becomes clear when we express the $C$-function as the integral over a disk of radius $r_{\star}$ on the plane
\begin{equation}\label{eq:planar-C-function}
c_{\text{pl}}(r_{\star}) := - 3\pi \int_{|z|=0}^{|z| = r_{\star}} d^2z |z|^2 \langle \Theta(z) \Theta(0) \rangle.
\end{equation}
We have replaced the double angle bracket $\langle \langle \rangle \rangle$ with the single bracket $\langle \rangle$ do denote that this quantity is now normalized by the partition function (it may diverge on the plane otherwise). It is convenient to denote \eqref{eq:planar-C-function} diagrammatically, as we do in Fig. \ref{fig:planar-c-func}.

The observation that string effective actions are intimately connected to $C$-functions is nothing new \cite{Ahmadain:2022tew,deAlwis-C-theorem-1988,TseytlinCentralCharge1987,Tseytlin:2006ak}. Indeed, they share a crucial common feature -- they are both stationary at $c=0$ CFTs. One key difference is that the planar $C$-function is stationary at \textit{all} CFTs, not just those with vanishing central charge. This stems from the fact that the $C$-function is computed on the plane, on which the coupling to the dilaton zero mode vanishes. The equation of motion of the dilaton zero mode is precisely what sets the constraint that $c=0$. Unfortunately, no $C$-theorem has been proven non-perturbatively on the sphere. \cite{TseytlinPerelmanEntropy2007} attempts to construct such a $C$-theorem with the help of Perelman's entropy functional (see \cite{papadopoulos2024scale,kim2024monotonicity} for recent discussions on this point), and is able to argue for a $C$-theorem to all orders in conformal perturbation theory within the \textbf{T2} renormalization window. 

\begin{figure}[ht]
\centering
\includegraphics[width=0.7\textwidth]{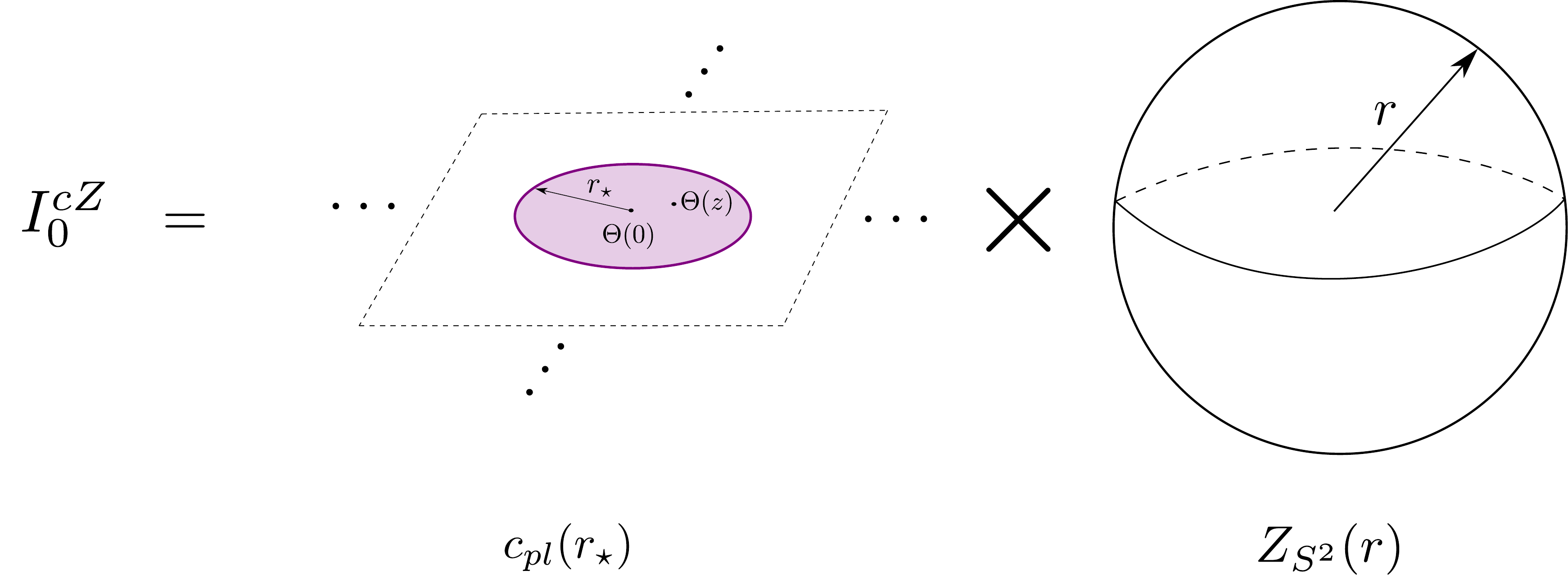}
\caption{A pictorial depiction of the planar cZ action. The $C$-function is drawn as in Fig. \ref{fig:planar-c-func}, and the sphere with no insertions represents $Z_{S^2}(r)$.}
\label{fig:cZ-diagram}
\end{figure}

We sidestep this issue by considering the product of the planar $C$-function $c_{\text{pl}}(r_{\star})$ and the sphere partition function $Z_{S^2}$:
\begin{equation}\label{eqn:plan-act-intro}
I_0^{cZ} = c_{\text{pl}}(r_{\star})Z_{S^2}.
\end{equation}
We also refer to it as the `$cZ$' action due to the form of its definition. The fact that \eqref{eqn:plan-act-intro} is a valid effective action non-perturbatively on the space of unitary 2d QFTs is the main result of this paper and is the subject of \S\ref{sec:non-pert}. We may again denote this product pictorially (see Fig. \ref{fig:cZ-diagram}).

Although valid non-perturbatively, this action is nonlocal in target space for two reasons. The first is that it is written as a product of two terms, each of which may be expressed as an integral over the zero modes of the worldsheet theory (as in e.g. \eqref{eqn:NLSM-Z0}). This nonlocality is not so bad at the classical level (i.e. the tree level we are working at), as the target space equations of motion only contain the $c_{\text{pl}}(r_{\star})$ factor. However, for very relevant worldsheet operators\footnote{Those with conformal dimension $\Delta \leq 1$}, $c_{\text{pl}}(r_{\star})$ is itself nonlocal in target space due to the noncompactness of the plane. This is because $n$-point functions pick up IR divergences as they are integrated over the plane. We discuss how these divergences are regulated non-perturbatively in \S\ref{ssec:cpt}. This leads to interesting nonanalytic behavior in the $C$-function, which we touch on in \S\ref{ssec:cpt} and \S\ref{sec:pert}. However, within the T2 renormalization window, we are able to find a field redefinition to $I_0^{\textbf{T2}}$.  Hence, within this window, the $cZ$ action is local up to field redefinitions.

We additionally perform a perturbative analysis of the $cZ$ action in \S\ref{sec:pert}, and in particular review that the planar $C$-function still contains the dilaton kinetic term despite not knowing about the dilaton zero mode. We present the entire infinite-dimensional mixing matrix of couplings for some simple minimally and non-minimally coupled operators, including primaries and simple descendants. We find that the $\Theta \Theta$ contact term can contribute terms to the $C$-function that are non-analytic in the coupling constants.

\subsection{Layout of Paper}

In \textbf{\S \ref{sec:review}} we review some essential technical background. This includes the Zamolodchikov $C$-theorem, conformal perturbation theory (in particular the issue of IR divergences). In \textbf{\S \ref{sec:non-pert}} we introduce the planar action and prove its validity on the space of unitary worldsheet theories (in Euclidean signature). We show that it is equivalent by field redefinition to the $I_0^{\textbf{T2}}$. Although most of our results are in Euclidean signature, we are able to generalize our results to a class of Lorentzian spacetimes (those which possess a timelike Killing field and are invariant under CTO, i.e.~the combination of charge-conjugation, time-reversal, and orientation-reversal.). In \textbf{\S \ref{sec:pert}} we analyze the planar action in perturbation theory and check its action on spurious tadpoles and irrelevant operators (including descendants). We demonstrate that all total derivative operators vanish from the effective action, in accordance with their status as vertex operators of pure gauge fields in the bulk. A similar interpretation is given to non-minimal couplings dressed with powers of the worldsheet Ricci curvature $R$ greater than 1. In \textbf{\S \ref{ssec:discussion}} we discuss our results and comment on future directions.

\section{Review}\label{sec:review}

For the reader's convenience, we review a few of the basic tools we use in this paper. This includes the Zamolodchikov $C$-theorem in \S\ref{ssec:c-thm} and conformal perturbation theory (and in particular the proper treatment of IR divergences) in \S\ref{ssec:cpt}.

\subsection{The Zamolodchikov $C$-theorem}\label{ssec:c-thm}

As we consider the RG flow of a unitary quantum field theory (QFT) into the IR, the degrees of freedom associated to massive fields are lost as these fields get gapped out. This loss of degrees of freedom renders RG flow irreversible. For unitary 2d QFTs on the plane, Zamolodchikov made this intuition precise by constructing a function of the coupling constants $c_{\text{pl}}(\lambda^a)$ that monotonically decreases under planar RG flow \cite{Zamolodchikov:1986gt}. It is defined through a combination of correlators of the stress-energy tensor
\begin{equation}\label{eqn:old-c-fnc}
\begin{split}
&C(|z|;\lambda^a):= z^4 \langle T(z)T(0) \rangle,\\
&F(|z|;\lambda^a):= z^3\bar{z} \langle T(z)\Theta(0) \rangle,\\
&H(|z|;\lambda^a):=z^2\bar{z}^2 \langle \Theta(z)\Theta(0)\rangle,\\
&c_{\text{pl}}(r_{\star};\lambda^a):= C(|z|=r_{\star};\lambda^a) - 2F(|z|=r_{\star};\lambda^a) - 3H(|z|=r_{\star};\lambda^a).
\end{split}
\end{equation}
For CFTs, $c_{\text{pl}}(r_{\star})$ evaluates to the central charge $c$, as we might expect from a quantity that tracks the number of degrees of freedom. Following \cite{Friedan:2009ik}, we may use the conservation laws to re-express $c_{\text{pl}}(r_{\star})$ as
\begin{equation}\label{eqn:int-c}
c_{\text{pl}}(r_{\star}) := - 3\pi \int_{|z|=0}^{|z| = r_{\star}} d^2z |z|^2 \langle \Theta(z) \Theta(0) \rangle.
\end{equation}
This is the two-point $\langle \Theta(z)\Theta(0) \rangle$ correlator integrated over a disc of radius $r_{\star}$ (see Fig. \ref{fig:planar-c-func}).

We immediately run into an subtlety -- the integrand in \eqref{eqn:int-c} may have arbitrarily sharp power law divergences at $z = 0$ when we allow for irrelevant deformations, whereas the expression defining $c_{\text{pl}}$ in \eqref{eqn:old-c-fnc} is manifestly finite as all correlators are evaluated at a finite distance. The catch is that the divergences from the $|z|=0$ limit in $\eqref{eqn:int-c}$ are infinite additive constants to $c_{\text{pl}}$ and have no bearing on its properties as a $C$-function as they do not contribute to the derivative of $c_{\text{pl}}$ along RG time. To make sense of $\eqref{eqn:int-c}$ we must therefore choose an appropriate renormalization scheme -- in particular, one that is valid on curved backgrounds. \cite{Ahmadain:2022eso} demonstrates the efficacy of heat-kernel regularization in the case of non-linear sigma models. In this paper we use dimensional regularization for general conformal perturbation theory\footnote{Occasionally, `conformal perturbation theory' is used to refer to deformations of a CFT only by marginal operators. This is \textit{not} how we use this term. We allow deformations by scalar operators of any dimension -- this is the sense in which conformal perturbation theory is used in e.g. \cite{Gaberdiel:2008fn,amoretti2017conformal}.}.

It is instructive to prove the $C$-theorem non-perturbatively starting from \eqref{eqn:int-c} directly. A small dilatation of the plane is equivalent to an infinitesimal increase of $r_{\star}$, so the variation of $c_{\text{pl}}(r_{\star})$ under RG flow is given by
\begin{equation}\label{eqn:c-dot}
-\frac{\partial}{\partial \log r_{\star}}c_{\text{pl}}(r_{\star}) = 3\pi r_{\star} \int_0^{2\pi} d\theta \langle \Theta(r_{\star}e^{i\theta})\Theta(0) \rangle = 6\pi^2 r_{\star} \langle \Theta(ir_{\star}/2) \Theta(-ir_{\star}/2) \rangle,
\end{equation}
where in the second equality we have assumed rotational and translational invariance. The points $z = \pm i r_{\star}/2$ are on the imaginary time axis, so we may write
\begin{equation}
\Theta(ir_{\star}/2) = e^{- H r_{\star}/2} \Theta(0) e^{H r_{\star}/2} = \Theta(-ir_{\star}/2)^{\dag},
\end{equation}
where $H$ is the worldsheet Hamiltonian. We recognize the expectation value on the RHS of \eqref{eqn:c-dot} as the norm squared of the state $\Theta(-ir_{\star}/2)\ket{0}$. Assuming the theory is unitary, this vanishes if and only if the theory is a CFT and is positive for all other reflection positive theories\footnote{Reflection positivity is the statement that in QFTs with a positive definite inner product, correlators symmetric about the Euclidean time axis must be positive.}. This proves
\begin{equation}
\begin{split}
&\dot{c}_{\text{pl}}(r_{\star}) \geq 0,\\
&\dot{c}_{\text{pl}}(r_{\star}) = 0 \iff \text{theory is a CFT}.
\end{split}
\end{equation}

We have proven that $c_{\text{pl}}(r_{\star})$ is stationary along RG flow for CFTs, but we have not yet proven that it is stationary with respect to all possible variations of the couplings. To do so, we only need to show that $c_{\text{pl}}(r_{\star})$ as defined in \eqref{eqn:int-c} is higher than first order in the coupling constants. To do so, recall that we may expand $\Theta(z) = \beta^a\mathcal{O}_a(z)$, so at finite separation we have
\begin{equation}
\langle \Theta(z) \Theta(0) \rangle = \beta^a \beta^b \langle \mathcal{O}_a(z) \mathcal{O}_b(0) \rangle.
\end{equation}
Therefore, at finite separation the integrand of \eqref{eqn:int-c} is at least second order in the couplings. When $z=0$, we may pick up contact terms from non-minimally coupled insertions:
\begin{equation}
\langle \Theta(0) \Theta(0) \rangle \sim \beta^{Ra} \langle \mathcal{O}_a \rangle. 
\end{equation}
Here, for emphasis, we use $\beta^{Ra}$ to specifically denote the beta function of the non-minimally coupled operator $R \mathcal{O}_a$. The Curci-Paffuti property \cite{curci1986consistency} guarantees that the beta function of non-minimally coupled operators vanishes at first order (we shall see in \S\ref{sec:pert} that there could be non-analytic contributions that are between first and second order).

To show $c_{\text{pl}}(r_{\star})$ indeed satisfies all of the criteria to be Zamolodchikov's $C$-function, we must demonstrate that it is equal to the central charge when the theory is a CFT. To do this, we note that CFTs satisfy
\begin{equation}
\langle \Theta(z) \Theta(0) \rangle = -\frac{c}{3\pi}\partial \bar{\partial} \delta^{(2)}(z) \implies -3\pi \int_0^{r_{\star}} d^2z |z|^2 \langle \Theta(z) \Theta(0) \rangle = c,
\end{equation}
so $c_{\text{pl}}(r_{\star})$ is simply equal to the central charge on CFTs.

\subsection{IR Divergences in Conformal Perturbation Theory}\label{ssec:cpt}

Perturbatively small fluctuations around classical string backgrounds are deformations of the $c=0$ CFT defining the background. In terms of the worldsheet action, these perturbations introduce an infinitesimal source $\lambda^a$ for some operator $\mathcal{O}_a$ in the theory:
\begin{equation}\label{eqn:S-pert}
    S_{QFT} = S_{CFT} + \int d^2z \sqrt{g(z)}\lambda^a\mathcal{O}_a(z).
\end{equation}
We treat such small deformations in the framework of conformal perturbation theory (CPT). We recommend \cite{amoretti2017conformal} for a review on the subject. If the operator $\mathcal{O}_a$ has dimension $\Delta$, $\lambda^{a}$ is fixed to have dimension $2 - \Delta$. Because the worldsheet is two-dimensional, operators with $\Delta > 2$ are irrelevent, $\Delta = 2$ are marginal, and $\Delta < 2$ are relevant. In the context of string theory, $\mathcal{O}_a(z)$ is the vertex operator of the corresponding target space field $\lambda^a$.

Given the perturbation to the action in \eqref{eqn:S-pert}, the action of the deformed theory is computed by an expectation value in the original CFT:
\begin{equation}
\begin{split}
&Z_{\text{QFT}} = \left\langle \exp(- \lambda^a\int \sqrt{g(z)}d^2z\, \mathcal{O}_a(z)) \right\rangle_{\text{CFT}} = \\
&=Z_{\text{CFT}} - \lambda^a \int \sqrt{g(z)}d^2z\ \left \langle \mathcal{O}_a(z) \right \rangle + \frac{1}{2}(\lambda^a)^2\int\sqrt{g(z)}d^2z_1d^2z_2 \left \langle \mathcal{O}_a(z_1)\mathcal{O}_a(z_2) \right \rangle + O(\lambda^3).
\end{split}
\end{equation}
All other expectation values in the theory are similarly computed as integrated correlators in the undeformed CFT:
\begin{equation}\label{eqn:A-pert}
\left \langle A(z) \right \rangle_{\text{QFT}} = \left \langle A(z) \right \rangle_{\text{CFT}} - \lambda^a \int \sqrt{g(z)}d^2w \left \langle A(z)\mathcal{O}_a(w) \right \rangle + O(\lambda^2).
\end{equation}
Given \eqref{eqn:int-c}, we will specifically be interested in computing perturbations to the $\Theta \Theta$ two-point function in \S\ref{sec:pert}.

The fact that the $\mathcal{O}_a(w)$ insertion in \eqref{eqn:A-pert} is integrated over the entire geometry naively leads to a problem for very relevant perturbations when the geometry has no IR cutoff (this is particularly relevant for computing the planar $C$-function). Suppose $\mathcal{O}_a$ is a primary $\mathcal{P}$ of dimension $\Delta < 1$, and we are computing the perturbation to the one-point function $\langle \mathcal{P}(0) \rangle_{\text{QFT}}$. The leading perturbation is the integrated two-point function:
\begin{equation}\label{eqn:pert-div}
\langle \mathcal{P}(0) \rangle_{\text{QFT}} = \frac{1}{2}\lambda^2 \int_{\mathbb{C}} d^2z \langle \mathcal{P}(z)\mathcal{P}(0) \rangle\, + O(\lambda^3) = \frac{1}{2}\lambda^2 \int_{\mathbb{C}} d^2z \frac{\kappa_{\mathcal{P}\mathcal{P}}}{|z|^{2\Delta}} + O(\lambda^3). 
\end{equation}
$\kappa_{\mathcal{P}\mathcal{P}}$ is just a constant normalizing the two-point function -- in particular it is the the relevant entry of the Zamolodchikov metric $\kappa_{ab}$. For all $\Delta \leq 1$, the right hand side is divergent. If we put in an explicit IR cutoff $\Lambda$, it diverges as $O(\Lambda^{2 - \Delta})$. Corrections with higher powers of $\lambda$ diverge even faster.

\begin{figure}[h]
\centering
\includegraphics[width=\textwidth]{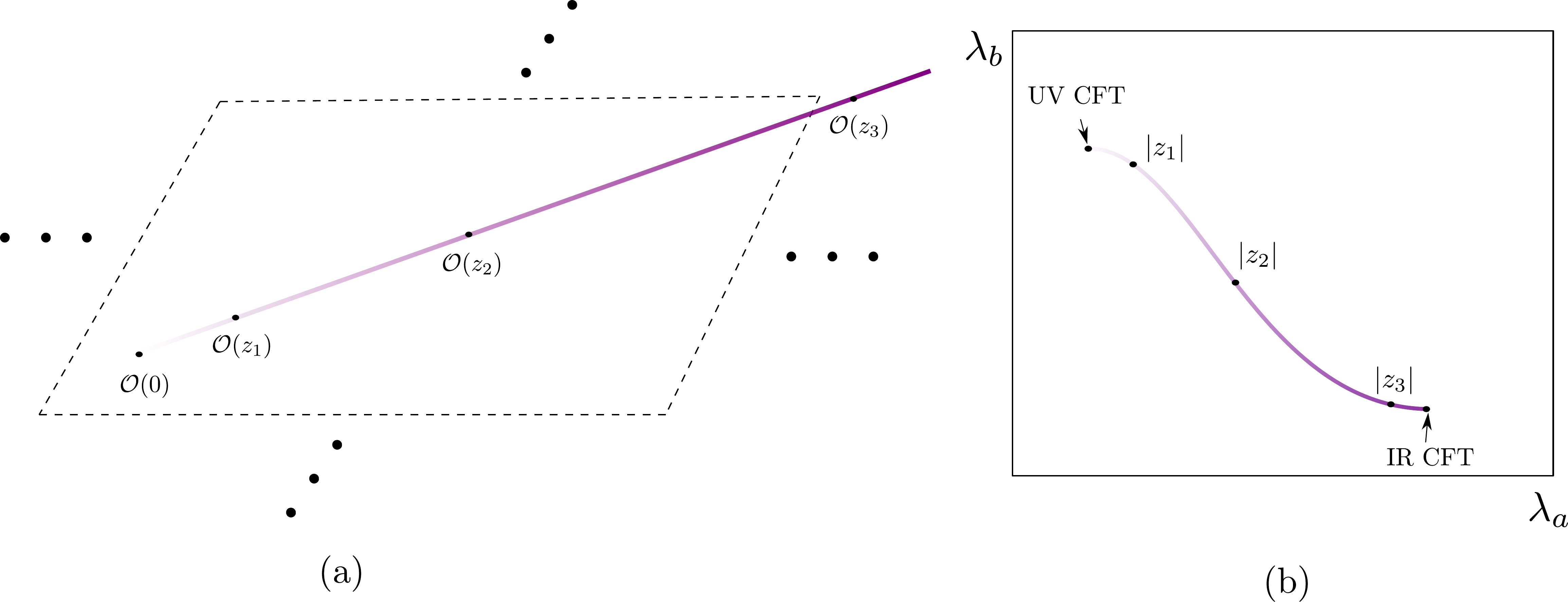}
\caption{In (a) we depict a two-point function $D(z):=\langle\mathcal{O}(z)\mathcal{O}(0)\rangle$ in a QFT deformed from a UV CFT by a relevant operator. A few points, $z_i$, are marked to denote some choices of length scale. In (b), we depict how increasing the length scale of the correlator corresponds to flowing from the UV CFT to the IR fixed point. Near $z_1$, $D(z)$ may be calculated in CPT with an irrelevant deformation of the UV CFT. At $z_2$, we are at large enough length scales that conformal perturbation theory has broken down. By the time we get to $z_3$, we have flowed so close to the IR CFT that the theory may be treated as a deformation by an \textit{irrelevant} perturbation of the IR CFT. Due to the properties of correlation functions of irrelevant deformations, this implies that at large enough length scales $D(z)$ falls off faster than $1/|z|^4$.}\label{fig:ir-flow}
\end{figure}

This puzzle arises precisely because $\mathcal{P}$ is a relevant operator. Considering ever-larger length scales is equivalent to flowing into the IR, so it is no surprise that the effect of relevant operators may grow as we increase the length scale. The resolution is to realize that at very long length scales we are no longer perturbatively close to the UV CFT we started at, and instead end up at an IR CFT deformed by an \textit{irrelevant} operator (see Fig. \ref{fig:ir-flow}). The two-point function of this operator will asymptotically fall off as $1/|z|^{\gamma}$ with $\gamma > 4$, and the length scale at which this behavior becomes dominant functions as an effective IR regulator that arises from non-perturbative effects.

For simple correlators, such as the one considered in \eqref{eqn:pert-div}, it is possible to extract the non-perturbative $\lambda$-dependence simply by dimensional analysis. $\langle \mathcal{P}(0) \rangle_{\text{QFT}}$ is the one-point function of an operator of dimension $\Delta$ in a theory where the only dimensionful parameter is its coupling $\lambda$. We may therefore deduce
\begin{equation}\label{eqn:non-anal}
\langle \mathcal{P}(0) \rangle_{\text{QFT}} = A_{\lambda} |\lambda|^{\Delta / (2 - \Delta)},
\end{equation}
where $A_{\lambda}$ is a dimensionless non-universal coefficient (that may indeed vanish, as it must for $\Delta > 2$, or depend on the sign of $\lambda$). This expectation value is no longer a smooth function of $\lambda$, and for $\Delta < 1$ in fact is the dominant contribution in the perturbative expansion. 

This regularization procedure works equally well if we are considering a two-point function of a relevant operator integrated against some function $f(z)$ that we may set by hand:
\begin{equation}\label{eqn:f-int}
    \langle \langle \mathcal{P} \mathcal{P}  \rangle \rangle_{f(z)} := \int d^2z\, f(z) \langle \mathcal{P}(z)\mathcal{P}(0) \rangle.
\end{equation}
So long as this integral converges for all irrelevant operators, the discussion above and the picture in Fig. \ref{fig:ir-flow} apply equally well and this integral will be rendered finite by non-perturbative effects. Because non-constant functions $f(z)$ must be defined in terms of additional dimensionful parameters, we may no longer use dimensional analysis to extract the dependence on the coupling constant as in \eqref{eqn:non-anal}. However, we may still introduce the effective length scale $\tilde{\Lambda}$ at which the $\langle \mathcal{P}(z) \mathcal{P}(0) \rangle$ is dominated by the IR CFT, and using both $\tilde{\Lambda}$ and the asymptotics of $f(z)$ we may bound the behavior of $\langle \langle \mathcal{P}(z) \mathcal{P}(0) \rangle \rangle_{f(z)}$. This will come particularly handy in \S\ref{ssec:field-redef}, where we consider $f(z) = \log |z|$.

\section{A Non-Perturbative Effective Action}\label{sec:non-pert}

Given a worldsheet QFT, the candidate for a non-perturbative effective action we consider is the product between the planar $C$-function $c_{\text{pl}}(r_{\star})$ and the sphere partition function $Z_{S^2}(r)$:
\begin{equation}\label{eqn:I0-prod}
I_0 = c(r_{\star})Z_{S^2}(r). \quad \quad (\text{The Planar $cZ$ Action})
\end{equation}
A diagrammatic version of this action was drawn in Fig. \ref{fig:cZ-diagram}. In this section we prove \eqref{eqn:I0-prod} works non-perturbatively on the space of unitary worldsheet theories, in the sense that its set of stationary points matches the set of on-shell string backgrounds. It is important to remember that $c_{\text{pl}}(r_{\star})$ is the sum of the $C$-function evaluated on the matter and ghost sectors separately, as they are decoupled. The ghost sector is always a $c=-26$ CFT, so if we label the matter $C$-function as $c_{\text{m},\text{pl}}(r_{\star})$ then $I_0$ takes the form
\begin{equation}
I_0 = \left(c_{\text{m},\text{pl}}(r_{\star}) - 26\right)Z_{S^2}.
\end{equation}

To prove \eqref{eqn:I0-prod} works, we note that a general variation of $I_0$ takes the form
\begin{equation}\label{eqn:gen-vary}
\delta I_0 = (\delta c_{\text{pl}}) Z_{S^2} + c_{\text{pl}}(\delta Z_{S^2}).
\end{equation}
In particular, the variation with respect to the dilaton zero mode is
\begin{equation} \label{eqn:dil-vary}
\frac{\partial I_0}{\partial \Phi_0} = -2I_0.
\end{equation}
First we consider the case $c_{\text{pl}} \neq 0$. In particular, this includes all CFTs with nonzero central charge. To show none of these worldsheet theories are stationary points, we note that for all unitary QFTs the sphere partition function $Z_{S^2}$ is nonzero as it computes the norm of the vacuum state. As a result the RHS of  \eqref{eqn:dil-vary} is nonzero iff $c_{\text{pl}}(r_{\star}) \neq 0$, so we do not have any stationary points when the $C$-function fails to vanish.

The only remaining task is to analyze variations around points in coupling space with $c_{\text{pl}} = 0$. If the theory is not a CFT on the plane, then we vary infinitesimally along RG flow time $t$. Zamolodchikov's $C$-theorem then guarantees
\begin{equation}
\delta_t I_0 = (\delta_t c_{\text{pl}})Z_{S^2} + c_{\text{pl}}(\delta_t Z_{S^2}) = (\delta_t c_{\text{pl}})Z_{S^2} < 0.
\end{equation}
Hence, for all theories with a nontrivial RG flow on the plane \eqref{eqn:I0-prod} is nonstationary.

Conversely, if the theory \textit{is} a $c = 0$ CFT on the plane, then for all variations both terms in \eqref{eqn:gen-vary} vanish. The first term vanishes by the c-theorem, and the second by the assumption that $c_{\text{pl}}=0$. We have therefore shown that a QFT is a fixed point of \eqref{eqn:I0-prod} if and only if it is a $c=0$ CFT on the plane. We have therefore accomplished what we set out to do -- find an action over the space of the entire set of 2d QFTs whose fixed points are precisely the set of $c=0$ CFTs.

Both $c_{\text{pl}}(r_{\star})$ and $Z_{S^2}(r)$ in \eqref{eqn:I0-prod} involve an integral over the zero modes of the worldsheet theory. For NLSMs, where we may separate the integral over the the zero modes $Y^{\mu}$ (see \cite{Ahmadain:2022eso,Tseytlin-SFT-EA-1986,TseytlinPerelmanEntropy2007,TseytlinZeroMode1989}), the planar action therefore takes the form 
\begin{equation}\label{eqn:I0-zero-modes}
I_0 = c_{\text{pl}}(r_{\star})\int d^DY^{\mu} Z_{S^2}(Y^{\mu};r).
\end{equation}
As $c_{\text{pl}}(r_{\star})$ itself involves an integral over $Y^{\mu}$, this action is inherently nonlocal in target space coordinates. However, precisely because $Z_{S^2} \propto e^{-2\Phi_0}$, the equations of motion are simply
\begin{equation}\label{eqn:c-eom}
c_{\text{pl}} = 0 \implies c_{\text{m},pl} = 26, \quad \forall \lambda^a\!:\; \frac{\delta c_{\text{pl}}}{\delta \lambda^a} = 0 \; .
\end{equation}
The particulars of $Z_{S^2}$ disappear from the equations of motion, which are local on the string scale and are simply the worldsheet $\beta$-functions. This suggests that there may exist a field redefinition off-shell that puts \eqref{eqn:I0-zero-modes} into a more local form. We propose such a redefiniton for the marignal and relevant sector of deformations in \S \ref{ssec:field-redef}.

\subsection{Gauging of Higher Nonminimal Couplings}\label{ssec:high-nonm}

We turn to the question of what happens to higher non-minimal couplings -- couplings of operators dressed with $R^p$ with $p \geq 2$. These are of particular interest, as it is these couplings that gave rise to the spurious tadpoles that limited the $I_0^{\textbf{T2}}$ regime of validity. The answer is that because we are working on the plane with only two $\Theta$ insertions, all higher non-minimal couplings completely vanish from $c_{\text{pl}}(r_{\star})$. As such, they vanish entirely from the equations of motion \eqref{eqn:c-eom}. For this reason, we interpret them as gauge redundancies.

To see this more explicitly, consider an action in terms of some variables $q_i,u_a,\phi_0$ that has the form \begin{equation}\label{eqn:ua-gauge}
S(q_i,u_a,\phi_0) = e^{-2\phi_0 - f(q_i,u_a)} \tilde{S}(q_i).
\end{equation}
We introduce two different indices, $i$ and $a$, to emphasize that there may be different numbers of $q$ and $u$ variables. The exponential factor, which depends on both the $q$ and $u$ variables, is analogous to $Z_{S^2}$. The $\tilde{S}$, which depends only on the $q_i$, is analogous to $c_{pl}$. The key property is that the dependence on $u_a$ is only through the exponential factor. This suggests a change of variables on $(\phi_0,u_a)$ that takes the form
\begin{equation}
(\phi_0, u_a, q_i) \rightarrow (\tilde{\phi}_0,u_a, q_i), \quad \tilde{\phi}_0 := \phi_0 + \frac{1}{2}f(u_a,q_i).
\end{equation}
The Jacobian of this transformation is just 1, so this is a valid change of coordinates. In these new coordinates, the action has the form
\begin{equation}
S(q_i,u_a,\tilde{\phi}_0) = e^{-2\tilde{\phi}_0}\tilde{S}(q_i),
\end{equation}
and is manifestly independent of the $u_a$, and so we may interpret these coordinates as pure gauge. In the $cZ$ action, the dependence of $I_0$ on the higher non-minimal couplings may be entirely absorbed into a redefinition of the dilaton zero mode, and therefore vanish entirely from the effective action. We may view $Z_{S^2}$ as an overall renormalization of the string coupling that is nonlocal in target space.

There is a particular facet of RG that makes the interpretation of higher nonminimal couplings as pure gauge particularly natural: a coupling constant of an operator dressed with $R^{p}$ cannot affect the RG flow of the coupling of an operator dressed with $R^{q}$, $q < p$. To see this, we again return to the relation $\Theta(z) = \beta^a \mathcal{O}(z)$. Consider a particular coupling $\lambda_R$ in some action of the form
\begin{equation}
S = \int d^2z\sqrt{g(z)}\left(\ldots + \lambda_R R^p \mathcal{O} + \ldots \right)
\end{equation}
Recall that if we write $\sqrt{g(z)} = 2e^{-2\omega}$, we may write $R = e^{2 \omega} \partial \bar{\partial} \omega$. The contribution to $\Theta$ will then be
\begin{equation}\label{eqn:lamb-r-beta}
\frac{\delta S}{\delta \omega(z)} = \ldots + 2p \lambda_R R^{p - 1} \partial \bar{\partial} \left(e^{-2\omega} \mathcal{O}\right) + \lambda_R R^p \delta_\omega \left( e^{-2 \omega} \mathcal{O}\right) + \ldots
\end{equation}
Total derivative terms cannot appear in the worldsheet action on a topology without boundary, so the first term cannot contribute to any beta functions. Therefore, $\lambda_R$ can only contribute to the beta functions of coupling constants dressed with at least $p$ powers of $R$. In this sense, quasi-minimal couplings may be viewed as a self-contained set of degrees of freedom whose equations of motion are unaffected by the higher non-minimal couplings.

Suppose we wish to consider the couplings of a particular theory that flows from some UV fixed point. We must deform this UV fixed point by a relevant operator to initiate the flow. In a unitary theory, such a relevant operator may only be a minimally coupled primary. Everywhere along this RG trajectory, the non-minimal couplings are uniquely determined by the minimal coupling that initiated the trajectory. To find a different theory with the same set of minimal couplings but a different configuration of higher non-minimal coupling, we would have to deform the original UV CFT by a purely higher non-minimal deformation. Such a deformation is purely irrelevant and will flow back to the initial UV CFT under RG, and so will not generate a new RG trajectory.

We therefore find an elegant option for gauge-fixing the gauged higher noniminimal couplings -- we restrict ourselves to the space of theories that flow from a UV fixed point. In particular, this includes all relevant deformations of any $c=0$ CFT, so all operators within the \textbf{T2} renormalization window are still allowed. It should be noted, however, that perturbatively near a $c=0$ CFT we may simply parameterize the space of all possible couplings with zero or one powers of $R$ (say some family of NLSMs), and set all higher non-minimal couplings to zero -- such a gauge fixing condition is much simpler than knowing which irrelevant deformations flow to a UV fixed point.

\subsection{Equivalence to T2 (Within the Renormalization Window)}\label{ssec:field-redef}

\begin{figure}[ht]
    \centering
    \includegraphics[width=\textwidth]{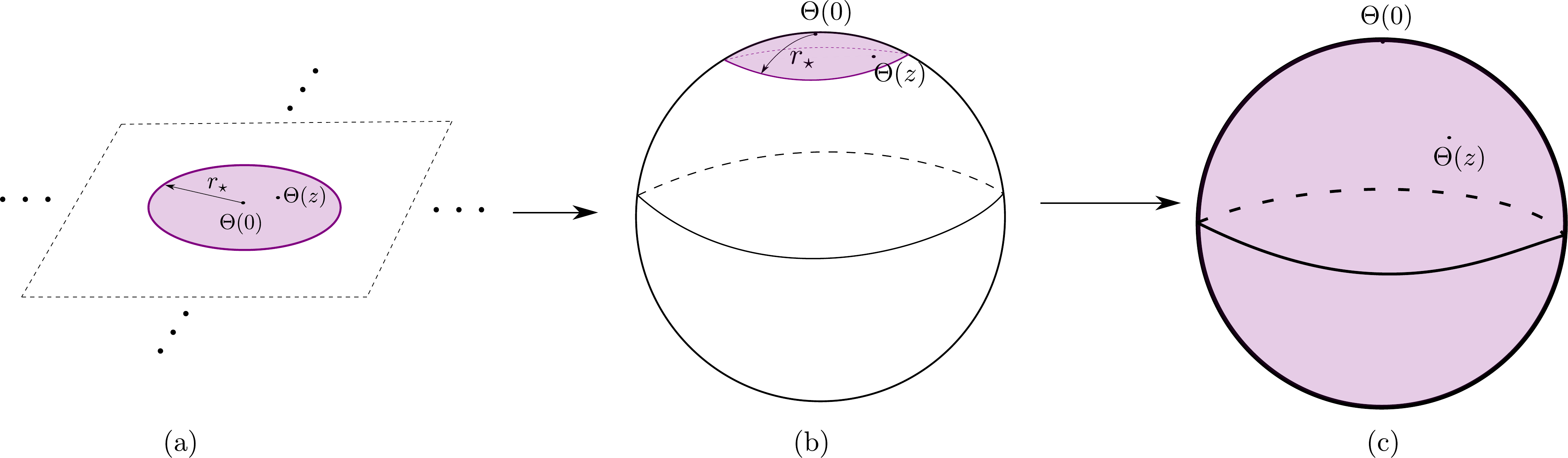}
    \caption{The pictorial form of the field redefinition from $c_{\text{pl}}(r_{\star})$ to \textbf{T2}. At each step, the shaded region denotes the disk-shaped region of integration of $\Theta(z)$. In (a) we start with the planar c-function as in Fig. \ref{fig:planar-c-func}. From (a) to (b), we perform the Weyl transformation \eqref{eqn:sphere-weyl} that deforms the plane to a very large sphere while keeping the radius of the disk $r_{\star}$ fixed. From (b) to (c), once the sphere is of finite radius, we add a term of the form \eqref{eqn:tt-redef} to fill in the rest of the sphere and bring it into the form of \textbf{T2}.}
    \label{fig:plane-to-T2}
\end{figure}

The authors of \cite{Ahmadain:2022tew} emphasize that while fixing the worldsheet Weyl frame seems to break worldsheet gauge invariance, off-shell physical quantities computed in different Weyl frames are related by field redefinition. We now review this argument, then show that \eqref{eqn:I0-prod} may be interpreted as a field redefinition of \textbf{T2} in the appropriate regime of validity. We do this by considering the process depicted in Fig. \ref{fig:plane-to-T2}, where we shrink the sphere\footnote{Considering the plane as a sphere of infinite radius.} that we compute $c_{\text{pl}}(r_{\star})$ on from infinite radius to finite radius while keeping the disk radius $r_{\star}$ finite. This corresponds to a non-uniform Weyl rescaling.

We now review the argument given in \cite{Ahmadain:2022tew}. Consider a (classical or quantum) system consisting of some collection of degrees of freedom $\phi^a$. These may be the generalized coordinates of some mechanical system, or the modes of some fields. For the compact target space cases we consider, we only need to worry about a countably infinite number of modes, so in any case the index $a$ is discrete. We consider an action on this space of modes, $S(\phi^a)$. A field redefinition is a coordinate transformation $\varphi^b(\phi^a)$ on the space parameterized by $\phi^a$. Perturbatively, we may write
\begin{equation}
\varphi^b \approx \phi^b + \delta \phi^b (\phi^a) + O\left((\delta \phi)^2\right).
\end{equation}
In a path integral, a generic field redefinition will introduce a Jacobian into the measure. At the classical level we are working at, we only have to worry about the effect on the action:
\begin{equation}\label{eqn:pert-qm-redef}
S(\phi^a + \delta \phi^a) = S(\phi^a) + E_a \delta \phi^a + O\left((\delta \phi)^2\right).
\end{equation}
We may read off from \eqref{eqn:pert-qm-redef} that an infinitesimal addition to $S$ proportional to the equations of motion $E_a$ may be absorbed into an infinitesimal field redefinition. If we work with a generic non-infinitesimal field redefinition $\varphi^b(\phi^a)$, we must integrate up the transformations \eqref{eqn:pert-qm-redef}. If at each point along the integration step we have independently proven we have a valid action (that has e.g. not introduced any spurious additional fixed points), we may be sure that our field redefinition is valid.

We now turn back to string theory. Consider some effective action $I$ computed on the space of 2d QFTs that depends on the choice of worldsheet Weyl parameter $\omega(z)$. We perturb $\omega(z) \rightarrow \omega(z) + \delta \omega(z)$. By definition of the beta functions (and because we are considering diffeomorphism-invariant theories on the worldsheet) $I$ will be perturbed as
\begin{equation}\label{eqn:I-dw-pert}
\delta I = \beta^a \int \sqrt{g(z)}d^2z \frac{\delta I}{\delta \lambda^a(z)}\delta \omega(z).
\end{equation}
If $I$ is a valid string effective action, the equations of motion $\beta^a$ are the equations of motion $E_a$ of the theory\footnote{To not clutter notation, we are not being careful with raising / lowering indices on the space of coupling constants. In principle, there should be a Zamolodchikov metric coming along for the ride.}, computed as
\begin{equation}\label{eqn:ba-Ea}
\beta^a = \frac{\partial I}{\partial \lambda^a} = E_a.
\end{equation}
We may now rewrite \eqref{eqn:I-dw-pert} as
\begin{equation}
\delta I = \frac{\partial I}{\partial \lambda^a} \int \sqrt{g(z)}d^2z \frac{\delta I}{\delta \lambda^a(z)}\delta \omega(z).
\end{equation}
We now note that this is precisely the same perturbation to $I$ we would find if we redefined the worldsheet coupling constants (i.e. the target space fields) by the infinitesimal perturbation
\begin{equation}\label{eqn:lamb-redef}
\delta \lambda^a = \int \sqrt{g(z)}d^2z \frac{\delta I}{\delta \lambda^a(z)}\delta \omega(z).
\end{equation}
This proves that an infinitesimal Weyl rescaling of the string effective action may be absorbed into a field redefinition. An amazing property of \eqref{eqn:lamb-redef} is that even though we allowed for a $z$-dependent Weyl rescaling $\delta \omega(z)$, the eventual redefinition of the coupling constants was $z$-independent. This means in order to make sense of a non-uniform Weyl transformation as a field redefinition, we do not have to introduce $z$-dependent coupling constants on the worldsheet as might naively be expected in the case of a local worldsheet RG flow. This is possible if and only if $I$ is a valid string effective action -- i.e. \eqref{eqn:ba-Ea} holds. This story applies equally well to an action of the form \eqref{eqn:I0-prod}, where we are only changing the geometry of one of the two factors in the product. The factor of $Z_{S^2}(r)$ simply comes along for the ride, and we arrive at the field redefinition
\begin{equation}\label{eqn:cZ-redef}
\delta \lambda^a = Z_{S^2}(r)\int \sqrt{g(z)}d^2z \frac{\delta c_{\text{pl}}(r_{\star})}{\delta \lambda^a(z)}\delta \omega(z).
\end{equation}

We now turn to the field redefinition induced by the Weyl transformation pictured Fig. \ref{fig:plane-to-T2}. We think of the plane as a sphere of infinite radius, then gradually shrink the radius until the the disk of radius $r_{\star}$ wraps around the sphere. As mentioned above, it is an important step to demonstrate that each action along the way is itself a valid action within the assumed renormalization window. This may be verified by direct computation, as the Weyl transformation properties of primary two-point functions may be computed straightforwardly as
\begin{equation}
\langle \mathcal{P}(z) \mathcal{P}(0) \rangle \rightarrow e^{-\Delta \omega(z)} \langle \mathcal{P}(z) \mathcal{P}(0) \rangle = e^{-\Delta \omega(z)}\frac{1}{|z|^{2\Delta}}.
\end{equation}
At each step in the field redefinition, the contribution from the deformation $\lambda \mathcal{P}(z)$ to the action is computed as
\begin{equation}\label{eqn:T2-interm}
I_0 \sim \int_0^{r_{\star}}d^2z\, e^{(2-\Delta) \omega(z)}\frac{(2-\Delta)^2}{|z|^{2\Delta-2}}.
\end{equation}
Note that this only vanishes if and only if $\Delta = 2$, so the action is valid.

We note that for unitary worldsheet theories, all higher non-minimal tadpoles are outside of the \textbf{T2} renormalization window at any order in perturbation theory. Furthermore, only marginal and relevant operators are allowed in the renormalization window to \textit{all} orders in perturbation theory, and this region of coupling space manifestly satisfies the constraint of flowing from a UV fixed point (said fixed point being the initial $c=0$ CFT we are perturbing around).

The second potential issue is that the initial perturbation $\delta \omega(z)$ we wish to consider takes us from the plane to some very large sphere of radius $K$, which corresponds to taking $\delta \omega(z)$ to be
\begin{equation}\label{eqn:sphere-weyl}
\delta \omega(z) = \log K^2 - \log(K^2 + |z|^2) \approx \begin{cases}
-\frac{|z|^2}{K^2}, \quad |z| \ll K,\\
-\log |z|^2 \quad |z| \gg K
\end{cases}.
\end{equation}
The `infinitesimal' perturbation $\delta \omega(z)$ gets arbitrarily large as $z$ goes to infinity. This is expected, as the plane is not infinitesimally close to a sphere of any radius.

This is not a problem, as we may show that for any worldsheet theory we may find a $K$ big enough such that the perturbation \eqref{eqn:cZ-redef} remains infinitesimal. In particular, we always consider $K \gg r_{\star}$. The problematic contribution begins at the point where $\delta \omega(z)$ is of order 1, which happens at $|z| \approx K$. The contribution to the RHS of \eqref{eqn:cZ-redef} is then
\begin{equation}
\begin{split}
&\delta \lambda^a_{\text{IR}} \sim Z_{S^2}(r) \int_{|z|=K}^{\infty} d^2z \frac{\delta c_{\text{pl}}}{\delta \lambda^a(z)}\log |z|^2\,=\\
&=Z_{S^2}\int_{|z|=K}^{\infty}d^2z \int_0^{r_{\star}}d^2w\,|w|^2\langle \Theta(w)\Theta(0)\mathcal{O}_a(z) \rangle \log|z|^2.
\end{split}
\end{equation}
We now employ the same trick as in \S\ref{ssec:cpt} (and \eqref{eqn:f-int} in particular) -- we consider large enough length scales $|z|$ such that the behavior of the three-point function in the integrand is dominated by the properties of the IR CFT\footnote{The assumption of the existence of an IR CFT is not necessary. We are just assuming that there is a non-perturbative effect that regularizes IR divergences, as in \S\ref{ssec:cpt}}. As before, call this length scale $\tilde{\Lambda}$. We are free to consider a sphere large enough such that $K \gg \tilde{\Lambda}$. Because the IR behavior of the theory is an irrelevant perturbation of the IR CFT, we have
\begin{equation}
    \int_0^{|w|=r_{\star}} d^2w\,|w|^2\langle \Theta(w)\Theta(0)\mathcal{O}_a(z) \rangle \sim \frac{C_{\lambda}}{|z|^{\gamma}}, \quad |z| \gg r_{\star},\tilde{\Lambda},
\end{equation}
for some exponent $\gamma > 4$ and overall constant $C_{\lambda}$. We then compute
\begin{equation}
    \delta \lambda_{IR}^a \sim Z_{S^2} \int_K^{\infty}d^2z \frac{C_{\lambda}}{|z|^{\gamma}} = 2\pi C_{\lambda}\frac{\log K}{(\gamma - 2)K^{\gamma - 2}}.
\end{equation}
As $\gamma > 4$, this contribution goes to 0 as $K$ goes to infinity.
    
So long as we take $K$ much larger than all relevant length scales, the field redefinition \eqref{eqn:cZ-redef} remains perturbative even when we make a topology-changing shift of Weyl frame from the plane to a sphere of radius $K$. We may then continue shrinking the sphere radius $K$ all the way down until $r_{\star}$ covers the entire sphere. To add the antipodal point (the south pole in Fig. \ref{fig:plane-to-T2}(c)), note that the two-point function
\begin{equation}
\langle \Theta(z)\Theta(0) \rangle  = \beta^a \beta^b \langle \mathcal{O}_a(z) \mathcal{O}_b(0) \rangle, 
\end{equation}
is proportional to the equations of motion. We may therefore add to the effective action $I_0$ any quantity of the form
\begin{equation}\label{eqn:tt-redef}
\delta I = \int_{\Sigma} \sqrt{g(z)}d^2z\, \langle \Theta(z) \Theta(0) \rangle = \beta^a \beta^b \int_{\Sigma}\sqrt{g(z)}d^2z\langle \mathcal{O}(z)\mathcal{O}(0) \rangle\,,
\end{equation}
where $\Sigma \subset S^2$ is any subregion of the sphere that does not include the point $z=0$ (to avoid contact terms). It is important to note that we can add this contribution bit by bit, so that the process depicted by the rightmost arrow in Fig. \ref{fig:plane-to-T2} gradually expands to cover the whole sphere instead of being added in all at once. At each point in this gradual process equation \eqref{eqn:T2-interm} applies, so we may again use the observation that at every point we have a valid effective action within the normalization condition.

We now return to the factor of $Z_{S^2}(r)$ that has been along for the ride in \eqref{eqn:cZ-redef}. When the process depicted in Fig. \ref{fig:plane-to-T2} causes the shaded disc to wrap the whole sphere, it is important to remember that the correlator we are instructed to compute is normalized (as it is written with a single pair of brackets $\langle \rangle$, unlike \eqref{eq:T2action-2pt-trace}). This simply means we are left with
\begin{equation}
I_0 = -Z_{S_2}(r) \int d^2z\,\sqrt{g(z)}|z|^2\langle \Theta(0) \Theta(z) \rangle_{QFT} = -\int d^2z\,\sqrt{g(z)}|z|^2\langle \langle \Theta(0) \Theta(z) \rangle \rangle_{QFT}.
\end{equation}
The role of $Z_{S^2}(r)$ is to turn the single bracket to a double bracket in the language of \cite{Ahmadain:2022tew} -- a normalized correlator into an unnormalized correlator. We have therefore arrived at the \textbf{T2} effective action via a continuous family of valid actions. This proves that $I_0^{\textbf{T2}}$ and the $cZ$ action are equivalent by field redefinition within the \textbf{T2} renormalization window. $cZ$, as a result, may be viewed as a direct extension of $I_0^{\textbf{T2}}$ to the space of massive target space modes.

It is important to understand how the \textbf{T2} prescription spoils the gauge symmetry discussed in \S\ref{ssec:high-nonm}. As can directly be seen from the spurious tadpoles of \textbf{T2}, higher nonminimal couplings appear nontrivially in the equations of motion computed from $I_0^{\textbf{T2}}$. Therefore, it is not true that $I_0^{\textbf{T2}}$ has the form \eqref{eqn:ua-gauge} with higher nonminimal couplings appearing only in the exponential. $I_0^{\textbf{T2}}$ \textit{does} have a family of gauged couplings -- those dressed with $(R - R_0)^p$ for $p \geq 2$ -- but they are not obviously related to the gauge directions of $cZ$ in any sensible way upon field redefinition.

\subsection{A Peek at Lorentzian Target Spaces}\label{ssec:CTO-Lor}

In principle we would like to generalize our results to Lorentzian target spaces. This is a challenge for a few reasons. First, even the simplest free Polyakov NLSM with a timelike scalar in its target space\footnote{Corresponding to strings propagating in a flat Minkowski spacetime} is non-unitary (in the sense that its Hilbert space contains negative-norm states). The $C$-theorem does not apply even perturbatively, and our proof fails. Second, the NLSM action with a timelike scalar is unbounded from below on any Euclidean worldsheet topology. We therefore need a prescription for analytically continuing the $X_0$ integral to a different contour, and it is not completely clear in general which prescription to choose\footnote{The situation is further complicated by the question of whether or not one should additionally analytically continue the worldsheet integration contour as well (see e.g. \cite{Witten:Feynman-Eps-StringTheory:2015,Schweigert:2001cu}).}.  Third, as the operator dimensions are now unbounded below, it is not clear that RG flow to the IR is well-defined, unless the spacetime is sufficiently smooth in the time direction.  Note that the Minkowksi vacuum is CTO symmetric for both bosonic and superstring theory\footnote{This is true even for heterotic superstrings, since while T and O considered separately each exchange the left- and right-movers, the combination TO does not.}.

However, we note that there is a specific class of Lorentzian target spaces for which the generalization of our effective action is particularly simple -- those that are both CTO invariant and possess a timelike Killing field\footnote{Spacetimes with such a field are called stationary. We avoid this term in the text in order to not conflate with the notion of stationary points of an action.}. Here, O flips orientation in target space -- equivalent to exchanging $z$ and $-\bar{z}$ on the worldsheet.  Hence, CTO in target space corresponds to a CPT transform of the worldsheet theory.  

The assumption of a timelike Killing field ensures that we may choose target space coordinates such that the NLSM couplings are independent of the timelike scalar $X_0$:
\begin{equation}\label{eqn:Killing-TS}
Z = \int \mathcal{D} X_i\, \exp(-S[X_i])\int \mathcal{D}X_0\, \exp(\int d^2z\, G_{00}(X_i) X^0 \partial \bar{\partial}X^0 + \ldots),
\end{equation}
where the ellipses denote an expansion in worldsheet operators that is analytic in the $X_0$ modes. The $X_0$ integral is therefore completely independent of the zero mode, and we may freely rotate the integration contour of the sphere zero from the Euclidean to the Lorentzian target space while only picking up a simple overall factor of $i$. We emphasize that we leave the sphere nonzero modes in Euclidean signature\footnote{The fact that the integral in \eqref{eqn:Killing-TS} is analytic in these modes motivates defining the Lorentzian integral by a rotation to the Euclidean contour. The idea behind keeping these modes on the Euclidean contour is that in the end, the effect of the target space Wick rotation is simply a single factor of $i$ in the action from the $X_0$ zero mode. This is precisely the behavior we want to make contact with a classical action describing gravity coupled to matter.}, where they behave as damped Gaussian integrals.

We may now prove that there are no spurious stationary points of the $cZ$ action that is invariant under these two symmetries (CTO and time translations). To see this, first note that all first order variations of the partition function that do not preserve the existence of a timelike Killing field vanish, as they correspond to computing the one-point function of a worldsheet operator that does not respect the symmetry of the NLSM. We may therefore restrict our analysis to the subspace of configuration space for which the decomposition \eqref{eqn:Killing-TS} applies, and there remains no subtlety in the choice of contour. We now use the fact that a CTO-invariant Lorentzian target space NLSM analytically continues to a reflection-positive NLSM with a Euclidean target space. As the property of being a $c=0$ CFT is preserved under such analytic continuation, if we find a CTO-invariant Lorentzian target space that is not a $c=0$ CFT but \textit{is} a stationary point of $I_0^{cZ}$, we may produce a reflection positive Euclidean target space that is also a spurious solution. This is impossible by the results of \S\ref{sec:non-pert}.

A perturbative proof of the validity of the $cZ$ action for more general Lorentzian target spaces would likely require a full analysis of all possible non-unitary Verma modules. A fully non-perturbative proof would likely require some novel idea for how to extend a $C$-theorem to 2d nonunitary QFTs. It would also be interesting to understand the relationship to the observation in \cite{Asano:2024edo} that the worldsheet theory with a Minkowski target space can be made unitary via a worldsheet wick rotation.

\section{Perturbative Analysis}\label{sec:pert}

While we have proven the validity of the action \eqref{eqn:I0-prod} non-perturbatively on the space of unitary QFTs, it is instructive to explicitly understand some of its perturbative features. In particular, we show how the planar $C$-function handles descendants and correctly reproduces the dilaton kinetic term. 

In this section we consider worldsheet theories of the form
\begin{equation}
\mathcal{L}_{QFT} = \mathcal{L}_{CFT} 
+ \mathcal{L}_{\text{pert}} = \mathcal{L}_{CFT} + \lambda^a\mathcal{O}_a + \tilde{\Phi}^a R \mathcal{O}_a, \quad S_{\text{pert}} := \int d^2z \mathcal{L}_{\text{pert}}.
\end{equation}
and expand perturbatively in the couplings $\lambda^a, \tilde{\Phi^a}$. Because the equations of motion \eqref{eqn:c-eom} only depend on the properties of the $C$-function, we need only perturbatively compute $c_{\text{pl}}(r_{\star})$ to study tree-level scattering. In conformal perturbation theory (see \S \ref{ssec:cpt}), the problem of computing the action reduces to computing integrated correlators. We denote by $\langle \rangle_{\text{CFT}}$ correlators in the ground state of the CFT we are perturbing around, and by $\langle \rangle_{\text{QFT}}$ correlators in the ground state of the perturbed theory. The $\Theta \Theta$ correlator becomes
\begin{equation}\label{eqn:CPT}
\begin{split}
&\langle \Theta(z) \Theta(0) \rangle_{\text{QFT}} = \frac{\delta}{\delta \omega(z)}\frac{\delta}{\delta \omega(0)}\left\langle \exp[-S_{\text{pert}}] \right \rangle_{\text{CFT}} =\\
&=\left\langle \exp[-S_{\text{pert}}]\left(\frac{\delta S_{\text{pert}}}{\delta \omega(z)}\frac{\delta S_{\text{pert}}}{\delta \omega(0)} - \frac{\delta^2 S_{\text{pert}}}{\delta \omega(z) \delta \omega(0)}\right) \right\rangle_{\text{CFT}},
\end{split}
\end{equation}
and a perturbative series in the coupling constants is formed simply by Taylor expanding the exponential. The second line contains a contact term, where both functional derivatives hit a single factor of $S_{\text{pert}}$. This contact term contains crucial physics, including accurately computing the central charge and dilaton kinetic term.

\subsection{Minimally Coupled Perturbations}

At leading order the perturbation to the off-diagonal stress tensor entries $T_{z \bar{z}}$ and $T_{\bar{z}z}$ is most easily determined from conservation laws (see Ch. 21 of \cite{Fradkin:2021zbi} for an instructive review):
\begin{equation}\label{eqn:no-bel}
\begin{split}
&\partial T_{z \bar{z}} = \bar{\partial}\int d^2w\,T(z)\mathcal{O}(w) = \bar{\partial}\int d^2w\, \sum_p \frac{L_p\mathcal{O}(w)}{(z-w)^{p+2}} = \frac{(-1)^p}{(p + 1)!}\partial^p L_p \mathcal{O}(z),\\
&\bar{\partial} T_{\bar{z}z} = \partial\int d^2w\,\bar{T}(z)\mathcal{O}(w) = \partial\int d^2w\, \sum_p \frac{\bar{L}_p\mathcal{O}(w)}{(z-w)^{p+2}} = \frac{(-1)^p}{(p + 1)!}\bar{\partial}^p \bar{L}_p \mathcal{O}(z).
\end{split}
\end{equation}
Note that for $p > 0$ these two expressions are in tension with each other, as these two expressions would imply that the stress tensor is not Hermitian (as $\bar{\partial} \partial T_{z\bar{z}} \neq \partial \bar{\partial} T_{\bar{z}z}$). Note that because $L_0 = \bar{L}_0$ for scalar deformations, $T_{z\bar{z}} - T_{\bar{z}z}$ is a total derivative. We may therefore restore conservation laws by adding an appropriate Belinfante improvement term of the form
\begin{equation}
T_{ab} \rightarrow T_{ab} + \partial_{c}\left(S^{abc} + S^{bac} - S^{cab}\right),
\end{equation}
where $S^{abc}$ is antisymmetric in $b$ and $c$. In two dimensions, $S$ only has four independent components which we may solve for to find
\begin{equation}\label{eqn:Thet-perp}
\Theta(z) = \left[(L_0 + \bar{L}_0 - 2) + 2\sum_{p \geq 1} \frac{1}{(p + 1)!}\left(L_{-1}^{p} L_p + \bar{L}_{-1}^{p} \bar{L}_p\right)\right]\lambda^a\mathcal{O}_a(z).
\end{equation}
If $\mathcal{O}_a$ is a primary $\mathcal{P}$, this simply reduces to the well known expression $\Theta(z) = (\Delta - 2)\lambda^a\mathcal{P}(z)$. The contribution to the $C$-function is therefore
\begin{equation}\label{eqn:hh-act}
c_{\text{pl}}(r_{\star}) = -3\pi {\lambda^a}^2 (\Delta - 2)^2\int_0^{r_{\star}}d^2z\,|z|^2 \frac{1}{|z|^{2\Delta}} = 3\pi {\lambda^a}^2 (\Delta - 2) r_{\star}^{4-2\Delta}.
\end{equation}
This is precisely the appropriate mass term for the target space field $\lambda^a(Y^{\mu})$ for all conformal dimensions $\Delta$. For all future such expressions we will drop the factors of $3 \pi$ and $r_{\star}^{4 - 2\Delta}$, which are again just constants that rescale the couplings. It is convenient to introduce the notation
\begin{equation}\label{eqn:L-thet}
L_{\Theta}:= (L_0 + \bar{L}_0 - 2) + 2\sum_{p \geq 1} \frac{1}{(p + 1)!}\left(L_{-1}^{p} L_p + \bar{L}_{-1}^{p} \bar{L}_p\right).
\end{equation}

It is further instructive to explicitly calculate the perturbation to $\Theta$ from an operator that is a total derivative -- $\mathcal{O}_a(z) = L_{-1}\mathcal{A}(z)$ for some $\mathcal{A}(z)$ of fixed scaling dimension ($\Delta - 1$). These terms may not contribute to the target space effective action as they are pure gauge in target space. Remembering $[L_{-1},L_p] = -(p+1)L_{p-1}$, we directly calculate
\begin{equation}
\begin{split}
&2\sum_{p \geq 1}\frac{1}{(p + 1)!}L_{-1}^pL_pL_{-1}\mathcal{A}(z) = 2\sum_{p \geq 1} \frac{1}{(p + 1)!}\left[L_{-1}^{p + 1}L_p - (p + 1)L_{-1}^pL_{p - 1}\right]\mathcal{A}(z) =\\
&= 2\sum_{p \geq 1} \frac{1}{(p + 1)!}L_{-1}^{p + 1}L_p - \sum_{q \geq 0}\frac{1}{(q + 1)!}L_{-1}^{q + 1}L_{q}\mathcal{A}(z) = -2L_{-1}L_{0}\mathcal{A}(z) = -(\Delta - 2)\L_{-1}\mathcal{A}(z).
\end{split}
\end{equation}
In the second equality, we have redefined the summation index to $q := p - 1$. The two sums cancel for $p \geq 1$, leaving only $-(\Delta - 2)L_{-1}\mathcal{A}(z)$, which precisely cancels the leading $(L_0 + \bar{L}_0 - 2)\mathcal{A}(z)$ term in \eqref{eqn:Thet-perp}. For any total holomorphic derivative perturbation only the $\bar{L}$ sum remains, leaving
\begin{equation}
\Theta(z) = 2\sum_{p \geq 1}\frac{1}{(p+1)!}\bar{L}_{-1}^p\bar{L}_pL_{-1}\mathcal{A}(z).
\end{equation}
This is a total divergence of the form $L_{-1}\bar{L}_{-1}\mathcal{B}(z)$, and may be removed from the action by an additional Belinfante improvement (equivalent to adding the non-minimally coupled perturbation $R \mathcal{B}(z)$ to the worldsheet Lagrangian density). We have therefore proven that all total derivative perturbations vanish from the target space effective action.

The most convenient way to analyze \eqref{eqn:Thet-perp} is to decompose Verma modules into total derivatives and special primaries. Special primaries are operators that are annihilated by $L_1$ -- the generator of special conformal transformations. While we use $\mathcal{P}$ to denote primaries, we use $\mathcal{SP}$ to denote special primaries. Invariance under special conformal transformations is enough to guarantee that only special primaries with the same conformal dimension may have a nontrivial two-point function. The simplest nontrivial example of such a special primary is
\begin{equation}\label{eqn:SP-level-2}
\mathcal{SP}_{-2}:=\left(L_{-2} - \frac{3}{2(2h+1)}L_{-1}^2\right)\mathcal{P},
\end{equation}
where the conformal weight of $\mathcal{P}$ is $(h,h+2)$ so that $\mathcal{SP}_{-2}$ is a scalar. We always define $\Delta$ to be the scaling dimension of the worldsheet deformation, so in this case we have $\Delta = 2h+4$. The two-point function of $L_{\Theta} \mathcal{SP}_{-2}$ is computed using \cite{Brehm:2020zri}\footnote{Specifically the Mathematica code given in Appendix B therein, as well as M. Headrick's Mathematica package Virasoro.nb (\hyperlink{https://sites.google.com/view/matthew-headrick/mathematica}{https://sites.google.com/view/matthew-headrick/mathematica).}}, and is given by
\begin{equation}\label{eqn:f1-def}
\begin{split}
&\langle L_{\Theta}\mathcal{SP}_{-2}(z) L_{\Theta}\mathcal{SP}_{-2}(0) \rangle := \frac{f_1(\Delta)}{|z|^{2\Delta}},\\
&f_1(\Delta)=\frac{(\Delta - 2)(32 \Delta^6 - 824 \Delta^5 + 8493 \Delta^4 - 44192 \Delta^3 - 120644 \Delta^2 - 159636 \Delta + 75024)}{18(\Delta - 3)}.
\end{split}
\end{equation}
Note that the zeros of $f_1(\Delta)$ occur at
\begin{equation}\label{eqn:sp2-f1-zeros}
f_1(\Delta) = 0 \implies \Delta \approx 6.96 \pm 1.59i\, ||\, \Delta = 4\, ||\, \Delta \approx 3.42 \pm 0.17i\, ||\, \Delta  = 2\, || \, \Delta \approx 0.99.
\end{equation}
Most zeros are complex. The zero at $\Delta = 2$ will be cancelled upon integration over the disk. There is a trivial zero at $\Delta = 4$, corresponding to the descendant of the identity. The zero at $\Delta \approx 0.99$ is precluded by the assumption of unitarity (remember, $\Delta$ is the conformal dimension of the descendant, not its primary ancestor).

\subsection{Non-minimal Couplings and the Mixing Matrix}

We now consider perturbations dressed with a factor of $R$. Given a particular special primary $\mathcal{SP}_a$, we may build an infinite chain of non-minimally coupled terms -- $R \partial^q \bar{\partial}^q \mathcal{SP}_a$ for any natural number $q$. We call the coupling to this operator $\tilde{\Phi}^{(q) a}$. For non-minimal couplings there are two terms we must worry about -- the first order perturbation to $\Theta(z)$ and the contact term. Let us deal with the former first. We may use $R(z) = -2\partial \bar{\partial}\omega(z)$ to write (absorbing the factor of 2 into a redefinition of $\tilde{\Phi}^{(q) a}$)
\begin{equation}
S_{\text{pert}} = -\tilde{\Phi}^{(q) a}\int d^2z\, \partial \bar{\partial} \omega(z) (\partial \bar{\partial})^2 \mathcal{SP}_a + O(\omega^2).
\end{equation}
The linear perturbation to $\Theta(z)$ is simply
\begin{equation}
\Theta(z) = -\tilde{\Phi}^{(q)a}(\partial \bar{\partial})^{q + 1}\mathcal{SP}_a + O(\tilde{\Phi}{^2}).
\end{equation}

For clarity and simplicity of the expressions, from this point we focus on cases where $\mathcal{SP}_a$ does not have a nontrivial two-point function with any other special primary (\eqref{eqn:SP-level-2} is an example). We additionally drop the $a$ index to remove clutter. In this case, the mixing matrix $M_{pq}$ takes the simple (albeit still $\infty \cross \infty$) form:
\begin{center}
\begin{tabular}{ c|c c c c c } 
& $\lambda$ & $\tilde{\Phi}^{(1)}$ & $\tilde{\Phi}^{(2)}$ & $\tilde{\Phi}^{(3)}$ & $\ldots$\\
 \hline
$\lambda$ &  $M_{00}$ & $M_{01}$ & $M_{02}$ & $M_{03}$ & $\ldots$ \\ 
$\tilde{\Phi}^{(1)}$ &  $M_{10}$ & $M_{11}$ & $M_{12}$ & $M_{13}$ & $\ldots$ \\ 
$\tilde{\Phi}^{(2)}$ &  $M_{20}$ & $M_{21}$ & $M_{22}$ & $M_{23}$ & $\ldots$ \\
$\tilde{\Phi}^{(3)}$ &  $M_{30}$ & $M_{31}$ & $M_{32}$ & $M_{33}$ & $\ldots$\\
$\vdots$ & $\vdots$ & $\vdots$ & $\vdots$ & $\vdots$ & $\ddots$
\end{tabular}
\end{center}
To compute all of the entries in this mixing matrix, we only need two correlators in addition to \eqref{eqn:f1-def}:
\begin{equation}
\langle L_{\Theta} \mathcal{SP}(z) L_{\Theta} \mathcal{SP}(0) \rangle:=\frac{f_1(\Delta)}{|z|^{2\Delta}}, \quad \langle \mathcal{SP}(z) L_{\Theta} \mathcal{SP}(0) \rangle:=\frac{f_2(\Delta)}{|z|^{2\Delta}}, \quad \langle \mathcal{SP}(z) \mathcal{SP}(0) \rangle:=\frac{f_3(\Delta)}{|z|^{2\Delta}}.
\end{equation}
Each of these terms is integrated against $|z|^2$ to find (we drop an overall normalization of $\frac{3}{2}\pi$):
\begin{align}\label{eqn:mix-mat-1}
&M_{00} = \frac{f_1(\Delta)}{(\Delta - 2)}r_{\star}^{2\Delta - 4},\\
&M_{q0} = M_{0q} = f_2(\Delta)\frac{\Gamma(\Delta + q)^2}{\Gamma(\Delta)^2(\Delta + q - 2)}r_{\star}^{2\Delta + 2q - 4} \quad q \geq 2,\\
&M_{pq} = f_3(\Delta)\frac{\Gamma(\Delta + p + q)^2}{\Gamma(\Delta)^2(\Delta + p + q 
- 2)}r_{\star}^{2\Delta + 2p + 2q - 4} \quad p,q \geq 1.
\end{align}

We have omitted $M_{01}$ for the moment due to the presence of a contact term. Contact terms are computed by taking
\begin{equation}
\frac{\delta^2 S_{\text{pert}}}{\delta \omega(z)\delta \omega(0)} = - \tilde{\Phi}^{(q) a} \partial^{q + 1}\bar{\partial}^{q + 1}\left[\delta(z) (L_{\Theta} + 2)\mathcal{SP}(z=0)\right].
\end{equation}
Note that the contact term is only nonzero inside a correlator if it hits an additional minimally coupled $\mathcal{SP}$ insertion. We then have
\begin{equation}
\begin{split}
&\lambda \tilde{\Phi}^{(q)} \int_{|z|^2=0}^{\infty} d^2z' \int d^2z |z|^2 (\partial \bar{\partial})^{q+1} \left[\delta(z) \left \langle \mathcal{SP}(z) \left(L_{\Theta} + 2\right) \mathcal{SP}(0) \right \rangle \right] =\\
&= \lambda \tilde{\Phi}^{(q)}\delta_{q,1}\left(f_2(\Delta) + 2 f_3(\Delta)\right)\int_{|z|^2=0}^{\infty} d^2z' \frac{1}{|z'|^{2\Delta}}.
\end{split}
\end{equation}
For $q \geq 1$ the contact term vanishes, as once we integrate the $\int d^2z$ integral by parts there are too many derivatives acting on $|z|^2$. For $q = 1$, this integral is divergent unless $f_2(\Delta) + 2f_3(\Delta)$ vanishes. If it is UV divergent, it may be removed by an appropriate renormalization condition. If it is IR divergent, we make sense of it using the non-perturbative, non-analytic behavior reviewed in \S\ref{ssec:cpt}:
\begin{equation}\label{eqn:non-ana}
\lambda\tilde{\Phi}^{(1)} \int_{|z|^2=0}^{\infty} d^2z \left \langle \mathcal{SP}(z) L_{\Theta} \mathcal{SP}(0) \right \rangle = A_{\mathcal{SP}}\tilde{\Phi}^{(1)}|\lambda|^{\Delta / (2 - 2\Delta)},
\end{equation}
for some undetermined non-universal coefficient $A_{\mathcal{SP}}$. Note that for unitary CFTs we have $\Delta > 0$ (aside from the trivial case of the identity operator), so despite the funny exponents in \eqref{eqn:non-ana} the $C$-function is higher than first order in the coupling constants and so is stationary.

We now consider a few simple examples of the functions $f_i(\Delta)$. For a primary, we have
\begin{equation}
f_1(\Delta) = \left(\Delta - 2\right)^2, \quad f_2(\Delta) = \Delta - 2, \quad f_3(\Delta) = 1.
\end{equation}
Note that $f_1(\Delta)$ precisely reproduces the $M_{00}$ term in \eqref{eqn:hh-act}. $M_{01}$ is particularly subtle in this case, and including the contact term takes the form
\begin{equation}\label{eqn:prim-M01}
M_{01} = (\Delta - 2)\Delta^2\int_{0}^{r_{\star}}d^2z\,\frac{1}{|z|^{2\Delta - 2}} + \Delta \int_0^{\infty}d^2z\,\frac{1}{|z|^{2\Delta - 2}}.
\end{equation}
Note that at $\Delta = 1$, the problematic UV logarithmic divergence is eliminated by a cancellation between the two terms in \eqref{eqn:prim-M01}, leaving us with an IR divergence that is again non-perturbatively regulated using the technology of \S\ref{ssec:cpt}. For non-primary deformations unitarity restricts us to $\Delta > 4$, where the contact term vanishes.

Let us return to the case of $\mathcal{SP}_{-2}$ as defined in \eqref{eqn:SP-level-2}. We have already computed $f_{1}(\Delta)$ in \eqref{eqn:f1-def}. We additionally have
\begin{equation}
\begin{split}
&f_2(\Delta):=\frac{16\Delta^4-214\Delta^3+919\Delta^2-1533\Delta+846}{4(\Delta-3)},\\
&f_3(\Delta):= \frac{9\Delta^4 - 99\Delta^3 + 374\Delta^2 - 539\Delta + 204}{18(\Delta-3)}.
\end{split}
\end{equation}
The corresponding zeros occur at
\begin{equation}
\begin{split}
&f_2(\Delta) = 0 \implies \Delta \approx 6.91\,||\,\Delta=4\,||\, \Delta \approx 3.31\,||\,\Delta=2\,||\,\Delta=1.16,\\
&f_3(\Delta) = 0 \implies \Delta=4\,||\, \Delta \approx 3.89\, || \, \Delta \approx 2.54\,||\, \Delta \approx 0.57.
\end{split}
\end{equation}

First note that $f_1(4) = f_2(4) = f_3(4) = 0$, implying that for descendants of the identity the mixing matrix is entirely 0. Now, we address the 0 in $f_2(\Delta)$ at $6.91$. Note that neither $f_1(\Delta)$ in \eqref{eqn:f1-def} nor $f_3(\Delta)$ have a 0 at this value, so the mixing matrix has the following form:
\begin{center}
\begin{tabular}{ c|c c c c c } 
& $\lambda$ & $\tilde{\Phi}^{(1)}$ & $\tilde{\Phi}^{(2)}$ & $\tilde{\Phi}^{(3)}$ & $\ldots$\\
 \hline
$\lambda$ &  $M_{00}$ & 0 & 0 & 0 & \ldots \\ 
$\tilde{\Phi}^{(1)}$ &  0 & $M_{11}$ & $M_{12}$ & $M_{13}$ & $\ldots$ \\ 
$\tilde{\Phi}^{(2)}$ &  0 & $M_{21}$ & $M_{22}$ & $M_{23}$ & $\ldots$ \\
$\tilde{\Phi}^{(3)}$ &  0 & $M_{31}$ & $M_{32}$ & $M_{33}$ & $\ldots$\\
$\vdots$ & $\vdots$ & $\vdots$ & $\vdots$ & $\vdots$ & $\ddots$
\end{tabular}
\end{center}
This makes it clear how the mixing matrix can still be nondegenerate at this value (as is guaranteed by the general non-perturbative properties of the $C$-theorem, as reviewed in \S\ref{ssec:c-thm}). Note further that neither $f_2(\Delta)$ nor $f_3(\Delta)$ vanish at the $\Delta \approx 0.99$ zero noted in \eqref{eqn:sp2-f1-zeros}, so the mixing matrix takes the form

\begin{center}
\begin{tabular}{ c|c c c c c } 
& $\lambda$ & $\tilde{\Phi}^{(1)}$ & $\tilde{\Phi}^{(2)}$ & $\tilde{\Phi}^{(3)}$ & $\ldots$\\
 \hline
$\lambda$ &  0 & $M_{01}$ & $M_{02}$ & $M_{03}$ & $\ldots$ \\ 
$\tilde{\Phi}^{(1)}$ &  $M_{10}$ & $M_{11}$ & $M_{12}$ & $M_{13}$ & $\ldots$ \\ 
$\tilde{\Phi}^{(2)}$ &  $M_{2 0}$ & $M_{21}$ & $M_{22}$ & $M_{23}$ & $\ldots$ \\
$\tilde{\Phi}^{(3)}$ &  $M_{30}$ & $M_{31}$ & $M_{32}$ & $M_{33}$ & $\ldots$\\
$\vdots$ & $\vdots$ & $\vdots$ & $\vdots$ & $\vdots$ & $\ddots$
\end{tabular}
\end{center}
Again, there are no obvious null directions. This gives promise that perhaps the validity of the $cZ$ action may be extended beyond the assumption of unitarity. We leave an explicit analysis of this possibility for future work.

\subsection{The Dilaton Kinetic Term}

Because $c_{\text{pl}}(r_{\star})$ is computed on a Ricci-flat topology,  it may be surprising that it correctly reproduces the dilaton kinetic term \cite{deAlwis-C-theorem-1988}. The clearest way to see how it arises is to consider a target space derivative expansion of the dilaton coupling around some point $X^{\mu} = Y^{\mu} + \eta^{\mu}(z)$:
\begin{equation}\label{eqn:dil-dev-exp}
R(z)\Phi(X(z)) = R(z)\left(\Phi(Y^{\mu}) + \eta^{\mu}(z)\partial_{\mu}\Phi(Y^{\mu}) + \ldots\right).
\end{equation}
The leading contribution to $c_{\text{pl}}(r_{\star})$ is then
\begin{equation}
\begin{split}
&\Theta(z) = \partial \bar{\partial} \eta^{\mu} \partial_{\mu}\Phi(Y^{\mu}) \implies \langle \Theta(z) \Theta(0) \rangle = \alpha' G^{\mu \nu}\partial_{\mu}\Phi \partial_{\nu}\Phi (\partial \bar{\partial})^2 \log|z| + O(\Phi, \alpha'),\\
&c_{\text{pl}}(r_{\star}) = -3\pi \alpha'G^{\mu \nu}\partial_{\mu}\Phi \partial_{\nu}\Phi \int_0^{r_{\star}}d^2z |z|^2 \partial \bar{\partial} \delta(z) = -3\pi \alpha'G^{\mu \nu}\partial_{\mu}\Phi \partial_{\nu}\Phi.
\end{split}
\end{equation}
This is exactly the form of kinetic term we are looking for. Going to higher orders in the Taylor expansion \eqref{eqn:dil-dev-exp} will lead to higher derivative terms for the dilaton, corresponding to $\alpha'$ corrections in the target space effective action.

More generally, the leading order terms in the action as you go off-shell are independent of field redefinitions.  It follows that their form is the same for $cZ$ at any value of $r_*$, and also \textbf{T2} at any value of $r$ (as long as we are within the renormalization window).  Since \textbf{T2} is local, it follows that---again only in the renormalization window---the leading order $cZ$ action also looks local in this regime.

\section{Discussion}\label{ssec:discussion}

In this work, we have generalized the results of \cite{Ahmadain:2022tew} beyond perturbation theory and beyond the scope of the \textbf{T2} renormalization window (or, indeed, any renormalization window). Our construction relied on a deep connection between $C$-functions and string effective actions, and in particular made use of the non-perturbative validity of the Zamolodchikov $C$-theorem on the plane. In our construction, higher non-minimal couplings are pure gauge. We have found a natural gauge-fixing condition is to restrict to irrelevant deformations that flow to a UV fixed point. Within the \textbf{T2} renormalization window (essentially the space of marginal and relevant deformations), we have demonstrated that our action is a field redefinition of the \textbf{T2} effective action. This implies that the $cZ$ action may be thought of as a direct extension of previous known effective actions, and in particular those computed in \cite{Tseytlin-SFT-EA-1986,Tseytlin:2006ak}.

We did not address higher-genus (string loop) corrections to the beta functions although Tseytlin's off-shell formalism covers $\textrm{g} > 0$ topologies \cite{TseytlinLoopCorrections1988,MTLoopCorrections1988,TseytlinLoopReview1989,TseytlinRGaStringLoops1990,Tseytlin-tadpole-div-1990} (reviewed in \cite{Ahmadain:2022tew}), as well as supersymmetric theories \cite{FT2,AT-PFOpenSuperstringEA1988,TseytlinGravitonAmlitudesEADisk1989}. Fishler and Susskind \cite {FischlerSusskind1:1986,FischlerSusskind2:1986} showed that the cosmological constant of closed bosonic strings can appear as a one-loop (genus) correction to the beta function for the background metric. This was further generalized to curved spacetime in \cite {Callan:1986bc,Callan:1988wz}. See also \cite{SeibergAnomalousDimension1986}. At least for the massless modes of the closed string, these works imply the existence of an off-shell string spacetime action whose minima are quantum-corrected beta functions. As noted in \cite{Callan:1986bc}, the main challenge would be to understand a genuinely new notion of generalized quantum conformal invariance and interpret the genus-corrected beta functions as equations of some generalized 2d NLSM worldsheet theory. Finding this NLSM theory would be a breakthrough but an even bigger one would be to include all couplings (not only the massless ones) as we did in this paper. We leave this for potential future work.

Additionally, we have identified a class of perturbations in \S\ref{ssec:high-nonm} that are pure gauge from the perspective of our effective action. To make contact with SFT, it is important to check that these operators are indeed projected out in the usual BRST formalism. It would also be interesting to further explore the relationship between the pure matter deformations to closed string backgrounds in closed SFT and their interpretation in the NLSM framework as second order shift in the dilaton \cite{Maccaferri:2023vns,Maccaferri:2024cjb}. In order to pursue a thorough study of quantum corrections to the classical action, we will need to carry out a full BRST analysis of our work to include operators with ghosts in our range of 2d NSLM couplings. As indicated in \cite{Callan:1986bc,Mansfield:1986it,Martinec:1986bq}, infinities at the quantum level break the off-shell tree-level BRST invariance, and it's only when we include all higher-genus worldsheet theories that we recover BRST symmetry. A study of the universal moduli space of Riemann surfaces and their boundaries will be required \cite{Friedan:1986ua,Friedan:1986hr}.

The endeavor to make string theory background-independent has gone hand in hand with properly treating off-shell target space backgrounds \cite{Witten-BSFT-Computations:1992,Sen-SFT-UV-Reg:2019,Sen:1990na}. It has consistently appeared that some form of worldsheet UV cutoff is central to the story. Moreover, it is precisely because the gravitational path integral treats all backgrounds and saddles on the same footing that it has surprisingly powerful insight into non-perturbative UV physics. The competition between saddles that are non-perturbatively far apart in field configuration space is directly responsible for the recent computation of the Page curve demonstrating unitary black hole evaporation from semiclassical gravity \cite{Penington:2019kki,Penington:2019npb,Almheiri:2019hni}. We hope that the program of developing a consistent effective action formalism for closed string theory, that may be manifestly described a patch-by-patch local target space action functional, may lead to important insights for how to bring known formulations of closed string field theory \cite{deLacroix:2017lif,Zwiebach:1992ie} closer to semiclassical physics. More broadly, many phenomena well described semiclassical gravity (such as real time evolution in the black hole interior) are only accessible through finite time evolution in target space, where boundary conditions and off-shell contributions are important and inescapable. We hope that further development of this line of work may shed light on the string-theoretic description of such physics.

\section*{Acknowledgements}

We would like to thank Minjae Cho, Theodore Erler, Luca Iliesiu, Rifath Khan, Manki Kim, Carlo Maccaferri, Raghu Mahajan, Prahar Mitra, Ronak M. Soni, and Xi Yin for insightful discussion. We are especially grateful to Zohar Komargodski, Savdeep Sethi, and Carlo Maccaferri for dedicating some of their time to review the manuscript. AF would especially like to thank Jan Boruch, Cynthia Yan, and Shunyu Yao for many lively conversations in Varian 361.

This research was supported in part by grant NSF PHY-2309135 to the Kavli Institute for Theoretical Physics (KITP).  In addition, AA is supported by The Royal Society and by STFC Consolidated Grant No. ST/X000648/1. AF is partially supported by the NSF GRFP under grant no. DGE-165-651.

AW was additionally supported by the AFOSR grant FA9550-19-1-0260 “Tensor Networks and Holographic Spacetime”, the STFC grant ST/P000681/1 “Particles, Fields and Extended Objects”, an Isaac Newton
Trust Early Career grant, and by NSF grant PHY-2207584 while working on this
paper during his sabbatical at the IAS.

{\footnotesize {\bf Open Access Statement} - For the purpose of open access, the authors have applied a Creative Commons Attribution (CC BY) licence to any Author Accepted Manuscript version arising.}

{\footnotesize{\textbf{Data access statement}: no new data were generated for this work.}}





\bibliography{refs.bib}


\end{document}